\documentclass[11pt,tightenlines,eqsecnum,floats,aps,amsmath,amssymb,nofootinbib,prd,shownopacs]{revtex4}

\usepackage{amsmath,amssymb,amsfonts,amsthm}
\usepackage{graphicx}
\usepackage{enumerate} 
\usepackage{color} 
\usepackage{subfigure}

\usepackage[mathscr]{eucal}
\usepackage[latin1]{inputenc}
\usepackage{amsmath}
\usepackage{amsfonts}
\usepackage{amssymb}
\usepackage{graphicx}
\usepackage{enumerate}
\def\a{\alpha}
\def\b{\beta}

\def\lp{{\ell}_{\rm Pl}}

\newcommand{\f}{\frac}

\newcommand{\fref}[1]{Fig.\,\ref{#1}}
\newcommand{\eref}[1]{eq.\,(\ref{#1})}

\def\f{\frac}

\def\mb{\bar \mu}

\usepackage{enumerate}

\newcommand{\be}{\nopagebreak[3]\begin{equation}}
\newcommand{\ee}{\end{equation}}
\newcommand{\bfig}{\nopagebreak[3]\begin{figure}}
\newcommand{\efig}{\end{figure}}
\newcommand{\ba}{\nopagebreak[3]\begin{eqnarray}}
\newcommand{\ea}{\end{eqnarray}}

\newcommand{\bmult}{\nopagebreak[3]\begin{multline}}
\newcommand{\emult}{\end{multline}}

\def\lp{{\ell}_{\rm Pl}}

\def\Heff{\mathcal{H}_{\rm eff}}
\def\Hmatt{\mathcal{H}_{\rm matt}}

\def\lp{l_{\rm Pl}}

\begin{document}

\title{Non-singular AdS-dS transitions in a landscape scenario}
\author{Brajesh Gupt}
\email{bgupt1@lsu.edu}

\author{Parampreet Singh}
\email{psingh@phys.lsu.edu}
\affiliation{Department of Physics and Astronomy, Louisiana State University, Baton Rouge, 70803}

\pacs{98.80.Cq, 04.60.Pp, 98.80.Qc}

\begin{abstract}
Understanding transitions between different vacua of a multiverse allowing eternal inflation is an open problem whose resolution is important to gain insights on the global structure of the spacetime as well as the problem of measure. In the classical theory, transitions from the anti-deSitter to deSitter vacua are forbidden due to the big crunch singularity. In this article, we consider toy landscape potentials: a double well and a triple well potential allowing anti-deSitter and de-Sitter vacua, in the effective dynamics of loop quantum cosmology for the $k=-1$ FRW model. We show that due to the non-perturbative quantum gravity effects as understood in loop quantum cosmology, non-singular anti-deSitter to de-Sitter transitions are possible. In the future evolution, an anti-deSitter bubble universe does not encounter a big crunch singularity but undergoes a big bounce occurring at a scale determined by the underlying quantum geometry. These non-singular transitions provide a mechanism through which a probe or a `watcher', used to define a local measure, can safely evolve through the bounce and geodesics can be smoothly extended from anti-deSitter to de-Sitter vacua.
\end{abstract}

\maketitle

\section{Introduction}

In the eternal inflation scenario based on new inflation \cite{stein,vil1}, the universe starts inflating
in a false vacuum producing a vast volume of the spacetime.  Different
parts of this vast spacetime inflate in different false vacua, giving rise to a landscape of
numerous vacua. At the scale of the entire vastness of the spacetime, inflation goes on
forever, producing an infinite number of pocket universes \cite{Guth:2007ng}.
 This picture of a big
universe leading to a fractal like structure of several pocket universes is often termed as a ``multiverse''. The idea of eternal inflationary multiverse has recently gained a lot of attention due to the possibility that it may naturally arise from the string landscape which is composed of many metastable vacua. Transitions of the configuration of the compact manifold in the string theory, results in transitions between different vacua in the 4-dimensional effective field theory description. A metastable vacuum with a positive minima leads to a de-Sitter (dS) evolution, and one with a negative minimum results in an anti-deSitter (AdS) phase of evolution in the landscape. A bubble universe which undergoes an AdS evolution reaches a maximum size determined by the local minimum of the effective potential, recollapses and evolves to a big crunch singularity in a finite time. Since a big crunch singularity is a generic property of AdS pocket universes in general relativity (GR), a transition from an AdS to a dS bubble is forbidden. It is therefore important to understand mechanisms originating from quantum gravity which can allow the transitions between AdS to dS vacua in the landscape to be non-singular.

The issue of obtaining such non-singular transitions is also important for a related problem of defining
appropriate measures for computing the probability of inflation to occur in an eternal inflationary multiverse (see Ref. \cite{Freivogel:2011eg} for a discussion of various measures used in this setting).
A promising proposal in the eternal inflation multiverse is to introduce a watcher, a probe which is conjectured to go through infinite transitions in the landscape \cite{Garriga:2012bc}. It is claimed that a local measure defined via such a watcher does not suffer from various ambiguities resulting from the choice of the cut-off surface.  Since the watcher encounters a big crunch singularity at the end of an AdS phase of evolution where the geodesics break down, the watcher proposal faces a severe limitation in the classical theory. This problem is also shared by other approaches to define a local measure in the multiverse \cite{m1,m2,m3,m4}. Unfortunately, current prescriptions to introduce a measure from a more fundamental theory in a landscape, such as using holographic ideas, also face difficulties associated with the big crunch singularity in the AdS evolution \cite{Garriga:2008ks,Garriga:2009hy}.

Since near the singularity,
the spacetime curvature becomes Planckian, the continuum spacetime description of general
relativity (GR) breaks down. It is expected that modifications to the dynamical equations originating from a quantum theory
of gravity would provide insights on the resolution of such singularities.
Although a full quantum
theory of gravity is yet to be found, loop quantum cosmology (LQC), a framework of homogeneous quantum
cosmology based on loop quantum gravity (LQG), presents a
promising avenue to tackle issues involving curvature singularities (see Ref. \cite{as1} for an extensive review).
 A key result in LQC for spatially flat isotropic spacetimes, is that non-perturbative quantum gravity effects originating from discrete quantum geometry in LQG cause a
bounce of the scale factor when the energy density of the universe reaches a universal maxima \cite{aps1,aps2,aps3,acs}. The big bang and big crunch singularities of the classical theory are replaced by a big bounce. The mechanism of non-singular bounce
in the isotropic spacetimes has been studied in great detail for several flat and curved models
with massless scalar field and other matter sources \cite{aps2,aps3,apsv,craigsingh13,ck13,acs,kv,szulck-1,bp,kp1,ap}.
Rigorous quantization of various anisotropic spacetimes has also been recently performed
\cite{chioub1,szulc_b1,awe2,awe3,we1,b1madrid1,b1madrid2,wemadrid1}.  It is notable that the bounce in loop quantized spacetimes occurs without
any fine tuning of the parameters or a specific choice of the energy conditions.

Evolution in LQC is governed by a quantum difference equation on the quantum geometry. For states which lead to a macroscopic universe peaked on a classical trajectory at late times, it is possible to obtain an effective continuum spacetime description \cite{jw,vt,psvt}.
The effective dynamics obtained from the effective Hamiltonian in LQC has been shown to be in excellent agreement with the underlying
quantum evolution for states which remain sharply peaked and bounce at scale factor larger than the Planck length \cite{psvt}. Based on the effective description of LQC, the strong singularities have been shown to be
resolved in isotropic \cite{ps09, sv} and Bianchi-I spacetime \cite{ps11}, and bounds on
energy density, shear scalar and expansion scalar in different models haven computed
\cite{cs09,gs1}. Given these results, LQC provides an excellent approach to address the difficulties, arising
due to big crunch singularity, to understand the transitions from the AdS to dS phase  in the effective 4-dimensional spacetime description of the multiverse.

Several investigations in the landscape paradigm imply that a pocket universe
is born out of a tunneling event from an other vacuum in the neighborhood.
The metric of the spacetime inside the bubble is given by
a $k=-1$ FRW universe using which several physical and observational consequences of bubbles have been studied in detail \cite{Gott:1982zf,Ratra:1994dm,Vilenkin:1996ar,Garriga:1998px,Barnard:2004qm,Freivogel:2005vv,Johnson:2011aa,Lehners:2012wz}, (see Refs. \cite{freiv,kleban} for recent reviews).
Our aim in this  this work is to use the insights from the new physics of non-perturbative quantum gravity as understood in LQC to understand transitions in multiple vacua of the landscape. We consider a $k=-1$  universe with a scalar field with self interacting potentials motivated by the landscape scenario in the multiverse in the effective spacetime description of LQC. We will  study the way loop quantum effects resolve the big crunch
singularity while making a non-singular transition from an AdS to a dS vacuum. 
It is important to emphasize some of the caveats in our analysis. We assume the validity of
effective equations in LQC for the $k=-1$ model. So far the quantization of $k=-1$ model in LQC is based on considering holonomies of the extrinsic curvature, unlike the Ashtekar-Barbero connection for spatially flat and positively curved spacetimes. It is possible that a quantization based on holonomies of connection may lead to some modifications to the effective equations, though these are not expected to change the qualitative aspects of dynamics. Further, strictly speaking, effective equations are valid only for scale factors greater than the Planck length \cite{psvt}. Thus, we focus our analysis on those cases where the bounce
occurs at scale factors much greater than the Planck length. On the same lines, given the limitations of the effective equations at scale factors comparable and smaller than the Planck length, any questions pertaining to such a regime can not be addressed in our approach.  Finally, we assume the existence of the self-interacting potentials motivated by the landscape scenario in LQC. Whether or not such scenarios can consistently arise in LQC at the level of the quantum theory is an open question, which goes beyond the scope of this work.\footnote{Our work will be on the lines of similar treatment of Ekpyrotic/Cyclic model in LQC, where singularity resolution was achieved assuming the existence of an effective 4-dimensional potential motivated from string theory \cite{cyc1,cyc2,cyc3}.}

The main result of this paper is based on the resolution of curvature singularities in LQC. Let
us consider a pocket universe described by a
negatively curved open FRW spacetime which is in an AdS vacuum at
some point in its evolution. In the classical description, such a pocket universe would
develop big crunch singularity in a finite proper time. In a striking contrast,  we show that
in LQC, the big crunch singularity is resolved due to the quantum geometric
corrections which become important in the Planck regime. As conjectured in the Ref.
\cite{Garriga:2012bc}, such a resolution of big crunch singularity would provide a way to
define a measure with respect to a `watcher' who travels on the worldline of the multiverse.
This in turn opens promising ways to address the measure problem in eternal inflation. We note that the goal of this work is neither to address the measure problem in the landscape scenario in LQC, nor the way quantum gravity effects may influence the probabilities for transitions.\footnote{For a discussion on the latter issue in chaotic inflationary scenario in LQC, see Ref. \cite{asloan_prob2,ck1}.} Our goal in this paper is to demonstrate that non-singular AdS-dS transitions can be achieved using loop quantum gravitational effects. 
Specifically, we show that depending on the initial conditions, the
pocket universe may transit to a dS vacuum in the future evolution or undergo another AdS phase before such a 
transition. This alters the global structure of multiverse, providing the watcher a safe passage from AdS to dS 
vacuum. 

It is to be  emphasized that a non-singular evolution of the scalar field from a negative part of the 
potential to a dS vacuum is a very non-trivial result. In the classical theory, such a transition can not be achieved due to singularity theorems and there is no well defined mechanism to obtain this, unless one considers violation of the weak energy condition. This problem has also been stressed in various earlier works (see for eg. \cite{Garriga:2012bc}), and it has been hoped that quantum gravity effects may resolve the big crunch singularity, and lead to a non-singular AdS-dS transitions. 
 Though in LQC singularities have been resolved in various models, one must exercise care while expecting results in the present setting. Here it is important to note that even though quantum gravity effects in LQC may resolve the singularity, it is not at all obvious that an AdS-dS transition would necessarily occur. An example on these lines occurs in the case of ekpyrotic model in isotropic LQC, where though the singularity is resolved, the scalar field {\it does not} make a transition from the negative to the positive part of the ekpyrotic potential \cite{cyc2}, unless one considers the presence of anisotropies \cite{cyc3}.\footnote{One should be  careful in a comparison of the landscape potential considered here with the analysis of the ekpyrotic potential such as in Ref. \cite{cyc2}. Since both the potentials have a negative part, one may be tempted to conclude that in the ekpyrotic potential one also has an AdS phase. However, as has been demonstrated in Ref. \cite{cyc2}, the approach to the singularity in ekpyrotic scenario is kinetic dominated, and thus there is no AdS phase near the big bang/crunch. Therefore, even though the landscape potentials (as considered in this work), and the ekpyrotic potential have some common features, there are sharp differences in the dynamical evolution and the equation of state on approach to the singularity.}  A closer example is the case of landscape potentials in $k=0$ isotropic cosmology (see Ref. \cite{piao} for an earlier work on these lines, however these results do not extend to the singularities encountered in the bubble universes as considered here.\footnote{Similar transitions from AdS to dS vacua has been discussed in the setting of cyclic inflation by considering quadratic corrections to the Friedmann equation in Ref. \cite{Biswas:2009fv,Biswas:2011qe}.}). As we show in the Appendix  \ref{app:k0}, though in the $k=0$ model, the big crunch singularity is resolved, there are no AdS-dS transitions. As in Ref. \cite{piao}, if the initial conditions are chosen in such a way that the energy density is positive in the negative vacua of the potential (which is possible by an appropriate choice of the kinetic energy of the scalar field), transitions from the negative to positive minima are obtained. However, we found that in such cases, the presence of the field in the negative minima does not correspond to an AdS phase of the evolution (see the Appendix \ref{app:k0}). For the spatially flat model, the form of the effective Friedmann equation is such that a negative energy density, which is required to capture the AdS like phase, leads to an imaginary Hubble rate.\footnote{In $k=0$ model, the Friedmann equation in LQC is: $H^2 = \f{8 \pi G}{3} \rho\left(1 - \f{\rho}{\rho_{\mathrm{crit}}}\right)$, where $\rho_{\mathrm{crit}} = 0.41 \rho_{\mathrm{Pl}}$.} Thus, the analysis of landscape motivated potentials in the spatially flat model is unable to capture the relevant physics of the AdS-dS transitions.

This paper is organized as follows. In the next section we briefly review the loop quantization
of an open FRW spacetime. We introduce the loop variables, the
SU(2) Ashtekar-Barbero connection $A_i^a$ and triads $E_a^i$, discuss their relation
with the usual metric variables and derive the dynamical equations both in the classical theory
and the effective spacetime description of LQC. In Sec. \ref{landscape}, we consider two different
types of landscape potentials: a double and a triple well potential, and solve the equations of motion
in these scenarios. We discuss the evolution of scale factor, scalar field and also provide
phase diagrams of the dynamical trajectories of the scalar field. In this section, we show
non-singular transition from AdS to dS vacua with help of numerical simulations. We
summarize and discuss our results in section \ref{discussion}.

\section{loop quantum cosmology of $k=-1$ FRW spacetime}
\label{lqc}
In this section we begin by an introduction of the FRW spacetime with negative spatial curvature in terms of
loop variables -- the Ashtekar-Barbero connection $A_a^i$ and triads $E_i^a$. We discuss
the relationship of these variables with usual metric variables, write the classical
Hamiltonian constraint and obtain the classical Friedmann and Raychaudhuri equations using the Hamilton's 
equations. Then, we discuss the effective description of loop quantum cosmology for
$k=-1$ spacetime.  The dynamical equations hence obtained are then used to derive
modified Friedmann equations for $k=-1$ FRW spacetime. We will also describe the
important features of the modified Friedmann equations which lead to vital differences
between the classical and effective LQC trajectories in the deep Planck regime.

\subsection{Classical theory}
We consider a homogeneous and isotropic open FRW spacetime with negative spatial
curvature, i.e. $k=-1$ with the following metric
\be
dS^2=-N^2 dt^2 + a^2(t)\left(\f{dr^2}{1-kr^2} + r^2 d\Omega^2\right),
\ee
where $N(t)$ is the lapse function, $a(t)$ is the scale factor, $r$ is the radial co-ordinate of
the spatial metric and $d\Omega^2$ is the metric on the surface of a 2-sphere. Thus the
spatial topology of the metric is $\Sigma=\mathbb{R}\times\mathbb{S}^2$. In order to define
symplectic structure on the spatial manifold we introduce a fiducial cell $\mathcal V$ whose
volume is given as ${\cal V}_o$. The edges of the fiducial cell are chosen to lie along the fiducial
triads $\mathring{e}_a^i$. The fiducial metric is $\mathring{q}_{ab}$ taken to be compatible
with the fiducial co-triads. Utilizing the underlying symmetry of $k=-1$ FRW spacetime the
Ashtekar variables can be written in terms of the symmetry reduced connection and triads as
follows
\be
A_a^i = c {\cal V}_o^{1/3} \mathring{\omega}, \qquad {\rm and} \qquad E_i^a= p~{\cal V}_o^{-2/3} \sqrt{q} \mathring{e}_i^a,
\ee
where $\mathring{e}_i^a$ are the densitized triads and $\mathring{\omega}_a^i$ are the
fiducial co-triads compatible with the fiducial metric $\mathring{q}_{ab}$. The symmetry
reduced connection, $c$ and the triad $p$ form a canonical pair which satisfies the
following Poisson bracket
\be
\{c,\,p\}=\f{8\pi G \gamma}{3},
\ee
where $\gamma \approx 0.2375$ is the Barbero-Immirizi parameter whose value of fixed via black hole
entropy computation in LQG. The triad, $p$ (whose orientation is chosen to be positive in this analysis) and the classical connection, $c$ are related to the scale
factor, $a$ and its time derivative as follows
\be
\label{eq:pca}p= a^2, \qquad c=\gamma \dot{a} + k,
\ee
where the `dot' represents the time derivative with respect to the proper time. It is important to note that the relation
between the connection, $c$ and the time derivative of the scale factor, as given above,
holds true only in the classical theory. This relation is modified in the effective description of LQC. On the other hand, the triad is given by the
same relation both in the classical and the effective description of LQC.

The classical Hamiltonian constraint, for FRW spacetime with the lapse function chosen to be
$N=1$, can be written in terms of the symmetry reduced connection and triad as follows
\be
\label{eq:classhamilt} \mathcal{H}_{\rm cl} = -\f{3}{8 \pi G \gamma^2} \sqrt{p} \Bigg[(c-k)^2 + k \gamma^2\Bigg] + \Hmatt.
\ee
From the vanishing of the Hamiltonian constraint, $\mathcal{H}_{\rm cl}=0$ we obtain
\be
\left(\f{c-k}{\gamma}\right)^2=\f{8\pi G}{3} \rho ~p - k
\ee
where $\rho$ is the energy density related to the matter part of the Hamiltonian as
$\Hmatt= \rho p^{3/2}$. Substituting the expressions of $p$ and $c$ in terms of metric
variables from \eref{eq:pca} to the above equation, one can obtain the classical Friedmann
equation
\be
H^2 = \left(\f{\dot a}{a}\right)^2= \f{8 \pi G}{3}\rho-\f{k}{a^2}.
\ee
The dynamical equations of the triad and the connection are given via the Hamilton's equation of motion
\be
\dot p = \{p,~\mathcal{H}_{\rm cl}\}, \qquad \dot c = \{c,~\mathcal{H}_{\rm cl}\}.
\ee
One can now use the equations of motion of the connection and the triad to obtain the
Raychaudhuri equation
\be
\dot H+ H^2=\f{\ddot a}{a} = -\f{4 \pi G}{3} \left(\rho + 3 P\right),
\ee
where $P$ denotes the pressure of the matter field. Using the classical Friedmann and Raychaudhuri equation, we can obtain classical trajectories for arbitrary matter. For the case of the positive cosmological constant, the future evolution asymptotes to dS phase. Where as for the negative cosmological constant, the future evolution is asymptotically AdS which ends in a big crunch singularity.

\subsection{Effective dynamics}
Quantization of a cosmological spacetimes in LQC is a symmetry reduced quantization based on the techniques of  LQG. The quantization procedure involves considering the field strength operator of the holonomies of the Ashtekar-Barbero connection, whose action on the states in the
geometrical representation leads to a quantum difference equation with uniform spacing in the volume. The quantum difference equation is a
direct consequence of the underlying quantum geometry, in particular of the minimum non-zero eigenvalue of the area operator in LQG. The evolution given by the quantum difference equation turns out
to be non-singular.
Due to certain technical difficulties, for the $k=-1$ model, one considers  field strength operator not constructed from the
holonomies of the connection, but of the extrinsic curvature, which results in a similar non-singular quantum difference equation.
Interestingly, for states which evolve to a macroscopic universe at late times, it is possible to derive an effective Hamiltonian constraint
using  geometrical formulation of quantum mechanics \cite{jw,vt,psvt}. Using  Hamilton's equation of
motion, effective dynamical equations are then derived from the effective Hamiltonian. These dynamical equations govern the evolution of a cosmological model under the
effective description of LQC, and in turn give rise to modified Friedmann and Raychaudhuri
equations. Various extensive numerical simulations show that the effective dynamical
trajectory stands in very good agreement with the full quantum evolution for sharply peaked
semi-classical states (see Ref. \cite{ps12} for a review).

The effective Hamiltonian for $k=-1$ FRW spacetime in terms of the symmetry reduced triad
($p$) and connection ($c$), with lapse ($N=1$),  given as \cite{kv,szulck-1,sv}
 \be\label{heff_k-1}
\Heff = -\f{3\, p^{3/2}}{8\pi  G \gamma^2 \lambda^2}\left(\sin\left(\mb \left(c -k\right)\right)^2 -k \chi\right) + \f{P_\phi^2}{2\,p^{3/2}} + V(\phi)\, p^{3/2},
 \ee
where $\mb = \lambda/\sqrt{|p|}$ is the length of the edge along which holonomy is
computed, $\lambda^2= 4\sqrt{3}\,\pi\, \gamma\, \lp^2$ is the minimum area gap determined by quantum geometry and
$\chi=-\gamma^2\mb^2$ for $k=-1$.
$P_\phi$ is the conjugate momentum of the scalar field and $V(\phi)$ is the self interacting
potential. It is useful to note that the above effective Hamiltonian is derived in the approximation when $p > 2.24 \lp^2$ \cite{sv}.\footnote{In conventions of Ref. \cite{sv}, this constraint corresponds to $v > 1$ for $\gamma \approx 0.2375$.} The dynamical equations for the connection and triad are given via the Hamilton's
equations of motion:
\be
\dot p=\{p,~\Heff\}, \qquad \dot c=\{c,~\Heff\},
\ee
which gives the following evolution equations for $c,~p,~P_\phi~\rm and~\phi$:
\ba
\label{cdot}\dot c &=& - \frac{8\pi\gamma}{3} \Bigg(\frac{3 p ^{3/2}}{8\pi  \gamma ^2 \lambda ^2} \left(\frac{\gamma ^2 \lambda ^2}{p ^2}-{\mb
 (1+c )} \cos\left({\mb (1+c )}\right) \sin\left({\mb  (1+c )}\right)\right) \nonumber \\ &&
  +\frac{9 \sqrt{p }}{16\pi  \gamma ^2 \lambda ^2} \left(-\frac{\gamma ^2 \lambda ^2}{p }+\sin\left({\mb  (1+c )}\right)^2\right) \nonumber \\ &&
  +\frac{3 P_\phi ^2}{4 p ^{5/2}}
-\frac{3}{2}\,  \sqrt{p }\, V(\phi)\Bigg)
\ea
\ba
\label{pdot}\dot p &=&\frac{2}{\gamma \lambda } \cos\left({\mb  (1+c )}\right) \sin\left({\mb  (1+c )}\right),\\
\label{phidot} \dot\phi  &=&\frac{P_\phi }{p ^{3/2}},\\
\label{pphidot} \dot P_\phi &=& -p^{3/2} \f{\partial V(\phi)}{\partial \phi}.
\ea
 Combining the dynamical equations for $\dot\phi$ and $\dot P_{\phi}$ it is straightforward to obtain the Klein-Gordon equation, which is equivalent to the conservation equation $\dot \rho = -3 H\left(\rho + P\right)$ (where $H = \dot p/4 p = \dot a/a$ is the Hubble rate):
 \be
\label{eq:kg}\ddot\phi + 3 H \dot \phi = -V_{,\phi}.
 \ee
The energy density and the pressure of the scalar field are given in terms of the velocity of
scalar field and potential as follows:
\be
\rho=\f{\dot\phi^2}{2} + V(\phi),~\qquad {\rm and} \qquad P=\f{\dot\phi^2}{2} - V(\phi).
\ee
The Klein-Gordon equation given via \eref{eq:kg} can be written in terms of the 
energy density and the pressure as follows:
\be
\label{eq:kgk-1}\dot{\rho}=-3H\left(\rho+P\right).
\ee
Using the dynamical equations hence obtained and the relation between the triad and
scale factor, we obtain the following modified Friedmann and Raychaudhuri equation for $k=-1$ in the effective description of LQC \cite{sv}:
 \ba
\label{friedk-1} H^2 &=& \left(\f{8\pi G}{3}\rho + \f{\mb^2}{\lambda^2}\right)\left(1-\f{\rho}{\rho_{\rm crit}} - \gamma^2\mb^2\right) \\
&=& \f{8\pi G}{3}\left(\rho+\rho_1\right)\f{1}{\rho_{\rm crit}}(\rho_2-\rho)
 \ea
 where $\rho_1=\f{3\mb^2}{8\pi G\lambda^2}$ and $\rho_2=\rho_{\rm crit}\left(1-\gamma^2\bar\mu^2\right)$. Note that both $\rho_1$ and $\rho_2$ are non-negative.\footnote{For $\rho_2$ to be negative, one requires $p < 0.2916 \lp^2$. However, the effective Hamiltonian (\ref{heff_k-1}) is valid only for  $p > 2.24 \lp^2$.}      
 \be
\label{raychaudh} \dot H = \left(-4\pi G\left(\rho + P\right) - \f{\xi + \gamma^2\mb^2}{\gamma^2\lambda^2}\right)\left(1-2\left(\f{\rho}{\rho_{\rm crit}}+\gamma^2\mb^2\right)\right)
 \ee
where $\xi = \sin^2(\mb) - \mb\sin(\mb)\cos(\mb)$ and $\rho_{\rm crit}=\f{3}{8\pi G\gamma^2\lambda^2}=0.41~\rho_{\rm Pl}$ is the upper bound on the energy density for isotropic models. The modified form of the Friedmann and Raychaudhuri equations lead to  resolution of various singularities which have been  studied in detail in the
Ref. \cite{sv}. The primary reason for the resolution of singularities lies in the quantum geometric effects which lead to the existence of $\rho_{\rm{crit}}$. Note that if quantum geometric effects are absent, i.e. if the limit $\lambda \rightarrow 0$ is taken, we recover classical Friedmann and Raychaudhuri equations. These equations are also recovered when the spacetime curvature is negligible compared to the Planck curvature.

From \eref{friedk-1} it is clear that for a physical solution, which requires $H^2\geq0$, the 
energy density satisfies the following inequality:
\be
-\rho_1\leq \rho \leq \rho_2.
\ee
This inequality clearly indicates that for the $k=-1$ FRW spacetime in LQC, the negative 
energy density is allowed as long as it remains greater than $-\rho_1$. Moreover, since 
$\rho$ can take positive values as well, the energy density can change sign and become 
positive during the evolution. Note that for the energy density to dynamically change sign from negative to positive, the time 
derivative of the energy density should be positive, when $\rho=0$ (at the point 
of sign change). It is straightforward to see that this is dynamically possible. If $\rho = 0$, then \eref{friedk-1} implies that $H^2 \neq 0$. Also, since $\rho = 0$ occurs for 
$\dot \phi^2/2 = - V(\phi)$, at the sign change of energy density, pressure is $P = \dot \phi^2 > 0$. Eq. (\ref{eq:kgk-1}), therefore implies that in the evolution when $\rho = 0$, $\dot \rho = - 3 H \dot \phi^2$. Thus, $\dot \rho > 0$ as the universe approaches a classical big crunch in the contracting branch after the recollapse. 
Hence, $k=-1$ FRW model not only allows negative energy density but also admits the 
change of sign of the energy density during the dynamical evolution. One can similarly carry out the above analysis for the $k=0$ case (see Appendix), and one finds that such a 
change of sign of energy density is dynamically forbidden in the spatially flat model. {\it Therefore,  unlike the flat model, the open FRW models can have transitions from  AdS (negative $\rho$) to dS (positive $\rho$) vacuum}.

\section{Landscape potentials and  `A\lowercase{d}S~\lowercase{to}~\lowercase{d}S' transitions}
\label{landscape}
In this section we consider two types of landscape potentials: (i) a double well
landscape potential with one AdS and one dS vacuum, and (ii) a triple well landscape potential with one AdS and
two dS vacua. We will consider these landscape potentials as self-interacting potentials of a
scalar field and study the evolution in the effective description of LQC.

\begin{figure}[tbh!]
\includegraphics[width=0.5\textwidth]{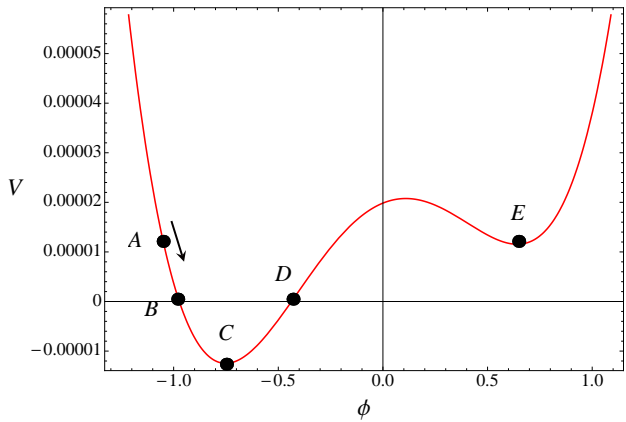}
\caption{A plot of the asymmetric double well potential motivated by the landscape scenario. Point `A' refers to the position of the scalar field for  one class of the initial conditions considered for evolution. Points `B' and `D' denote the zeros of the potential, `C' marks the AdS vacuum and `E' corresponds to the dS vacuum where the value of the potential is positive and locally minimum.}
\label{fig:potential}
\end{figure}

\subsection{Double well potential}
Let us consider a landscape potential as shown in \fref{fig:potential}. In order to model such a potential, we consider the following polynomial expression
\be
\label{pot}V(\phi)= V_o \left(\left(\f{\phi^2}{\a^2}-\delta\right)^2 + \f{\b}{\a}\phi\right),
\ee
where the parameters $\a,~\b ~ {\rm and}~\delta$ determine the shape of the potential;
$\a$ controls the horizontal distance between the two local minima of the potential, $\b$
governs the vertical distance between the two local minima and $\delta$ governs the
position of the $V|_{\phi=0}$ for given values of $\a\,{\rm and}\,\b$.

The point `A' in \fref{fig:potential} corresponds to a point in the positive part of the potential
where one set of initial conditions are given in our analysis. The arrow in \fref{fig:potential}
denotes that the field is considered to be initially rolling down, and enters the AdS well in future
evolution. We also consider the initial conditions on the field such that the field begins the
evolution at the bottom of the AdS well denoted by `C'. The points `B' and `D' denote the two zeros of
the potential and `E' marks the position of the positive local minimum. The vacuum `C' of the
potential corresponds to negative potential and would give rise to negative energy density. If the
field spends some time in the minimum of the potential, then it mimics
an AdS like evolution with almost constant negative energy density. This vacuum is referred to
as the AdS vacuum. The second vacuum, denoted by `D' has a positive potential which would
mimic a positive cosmological constant (dS) vacuum. If the field is stuck in the positive side of
the potential with a small field momentum, the bubble universe undergoes a de-Sitter (dS) like
evolution. The value of the ``effective cosmological constant" in this model can be controlled by
changing the value of the magnitude of the potential $V_o$. The dynamics of the field in this
potential has interesting phenomenological features due to AdS and dS like phases.
In the classical theory, the evolution in AdS phase ends with a  big-crunch singularity. Therefore, once the field comes into the negative
potential regime, the spacetime has no escape but to collapse into big-crunch. In contrast, in LQC due to underlying quantum geometric effects, the big-crunch type
singularity is avoided and evolution continues to an expanding branch.

The matter Hamiltonian for the scalar field with the potential \eref{pot}, is given as
\be
\Hmatt = \f{P_\phi^2}{2\, p^{3/2}} + p^{3/2}\,V_o \left(\left(\f{\phi^2}{\a^2}-\delta\right)^2 + \f{\b}{\a}\phi\right).
\ee
In the following, we  solve the effective dynamical equations for the above matter
Hamiltonian. The set of equations (\ref{cdot}-\ref{pphidot}) form a well posed initial value
problem. That is, by providing initial conditions on ($p,~c,~\phi,~P_\phi$) at a given point of
time $t=t_o$, the state of universe at a later time $t$ can be computed. Since, the initial
conditions must satisfy the effective Hamiltonian constraint, we provide the initial values of
three of the variables and calculate the initial value of the remaining one from the
vanishing of the effective Hamiltonian constraint, $\Heff\approx0$. Typically we provide
$p(0),~\phi(0),~{\rm and}~P_\phi(0)$ and calculate $c(0)$ from the effective Hamiltonian
constraint. We investigate the numerical evolution by providing the initial conditions on the scalar
field at two different places in the landscape potential. In the first, the field is considered to roll
down from a positive part of the potential (corresponding to point `A'), and in the second, the
field starts to evolve from the bottom of the AdS well (corresponding to point `C'). The form of the dynamical equations are
too complicated to be solved analytically, therefore we will solve them numerically.

\subsubsection{Evolution in double well potential: initial conditions at `A'}
\begin{figure}[tbh!]
    \subfigure[]
    {
    \includegraphics[width=0.47\textwidth]{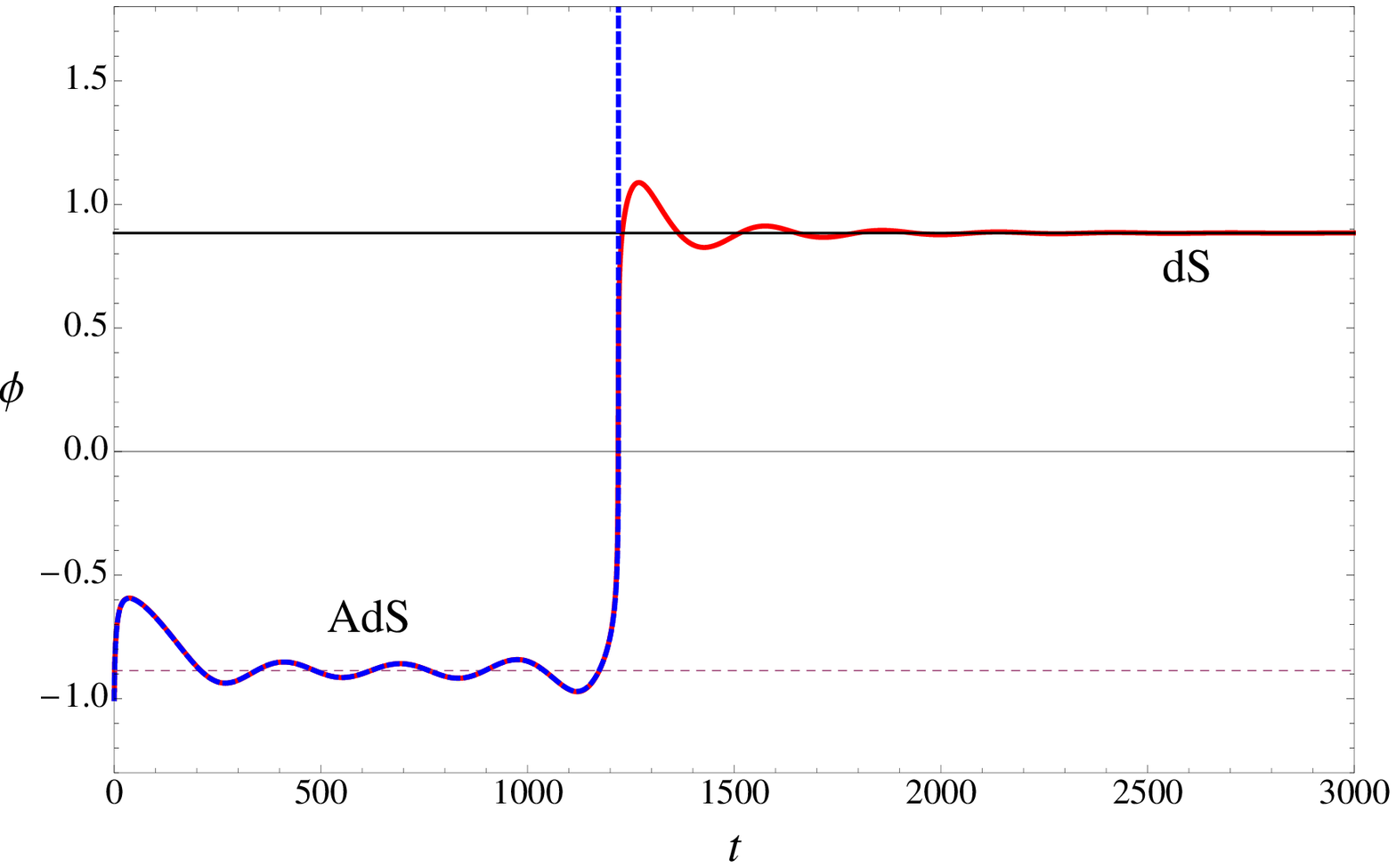}
    \label{fig:phidw1}
    }
    \subfigure[]
    {
    \includegraphics[width=0.475\textwidth]{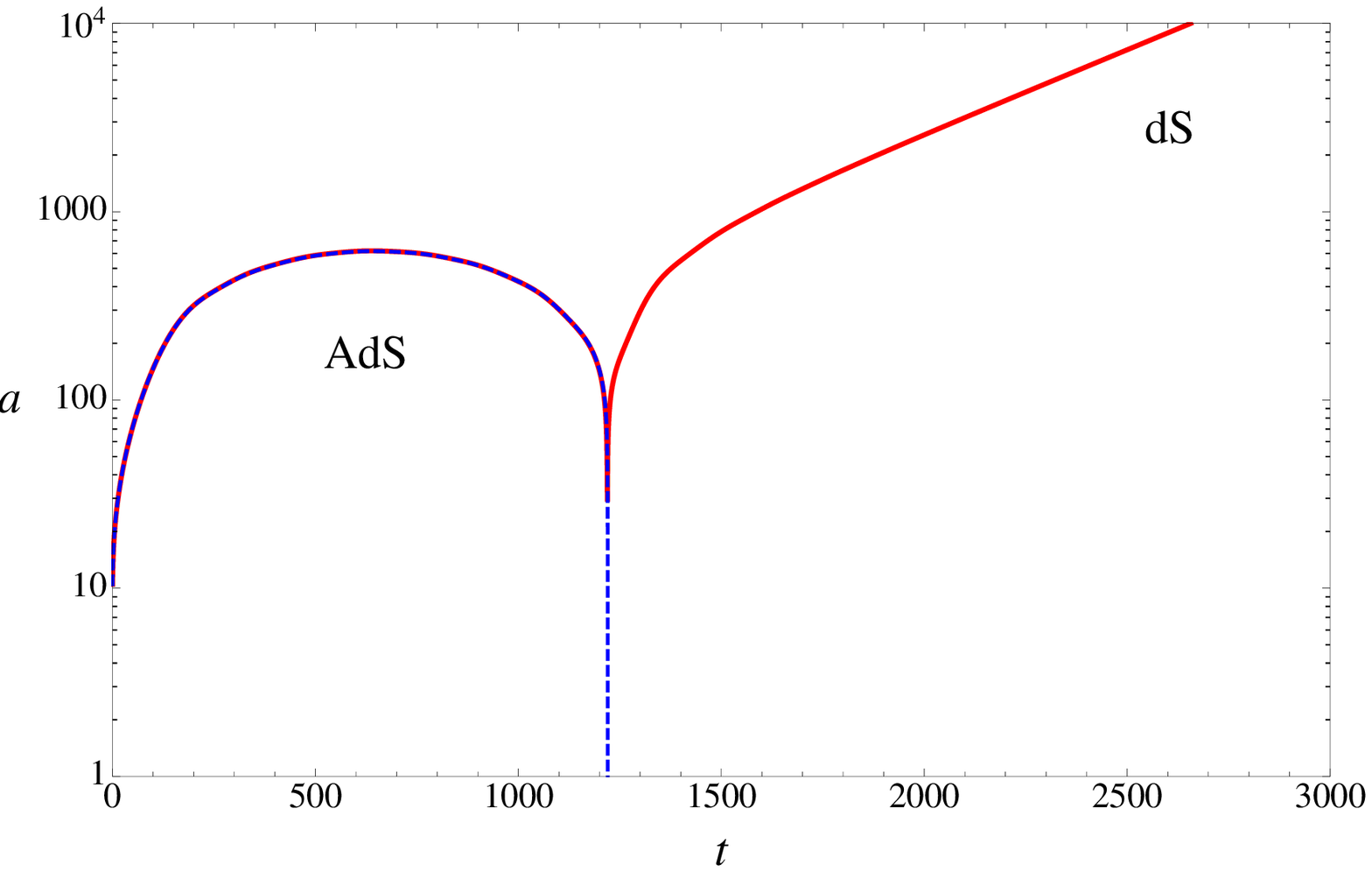}
    \label{fig:adw1}
    }
 \caption{These figures show the evolution starting from the left of the AdS vacuum of the
 landscape shown in \fref{fig:potential}. Figures (a) and (b) respectively show the evolution
 of the scalar field and the scale factor. The dashed curves show the classical trajectory, while the solid curves correspond to the LQC trajectories. It is
 clear from these figures that in LQC, instead of big-crunch singularity, there is a
 non-singular bounce. Following the bounce universe transits to a dS phase as the field
 evolves to the dS vacuum. The initial conditions for these plots are:
 $a(0)=10.5,~\phi(0)=-1.4,~\dot\phi(0)=0.14$ (in Planck units). The parameters of the
 potential are taken to be: $\alpha=0.28,~\beta=0.16,~ \delta = 10$ and $V_o = 10^{-6}$.}
  \label{fig:doublewell1}
\end{figure}

\begin{figure}[tbh!]
    \subfigure[]
    {
    \includegraphics[width=0.49\textwidth]{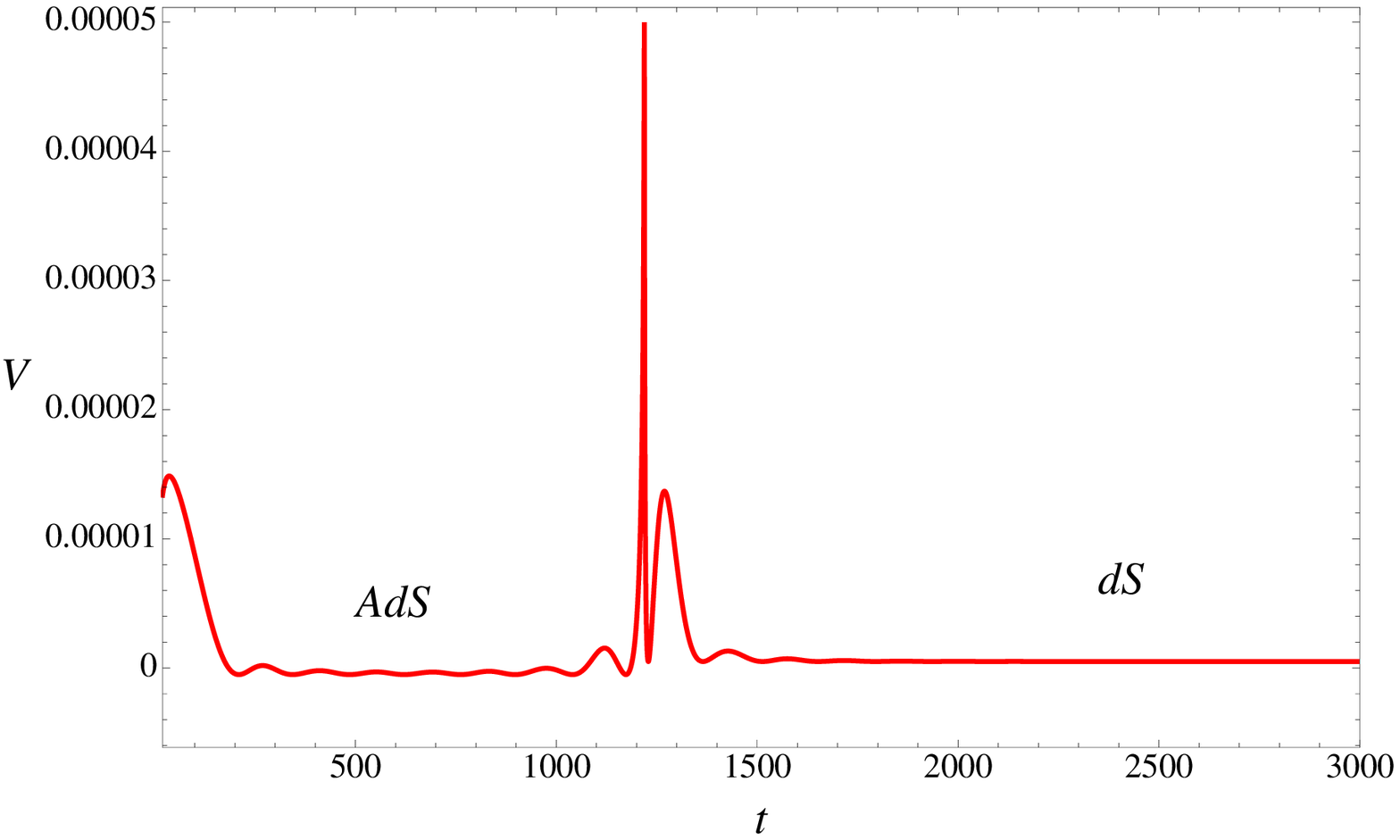}
    \label{fig:potentialdw1}
    }
    \subfigure[]
    {
    \includegraphics[width=0.475\textwidth]{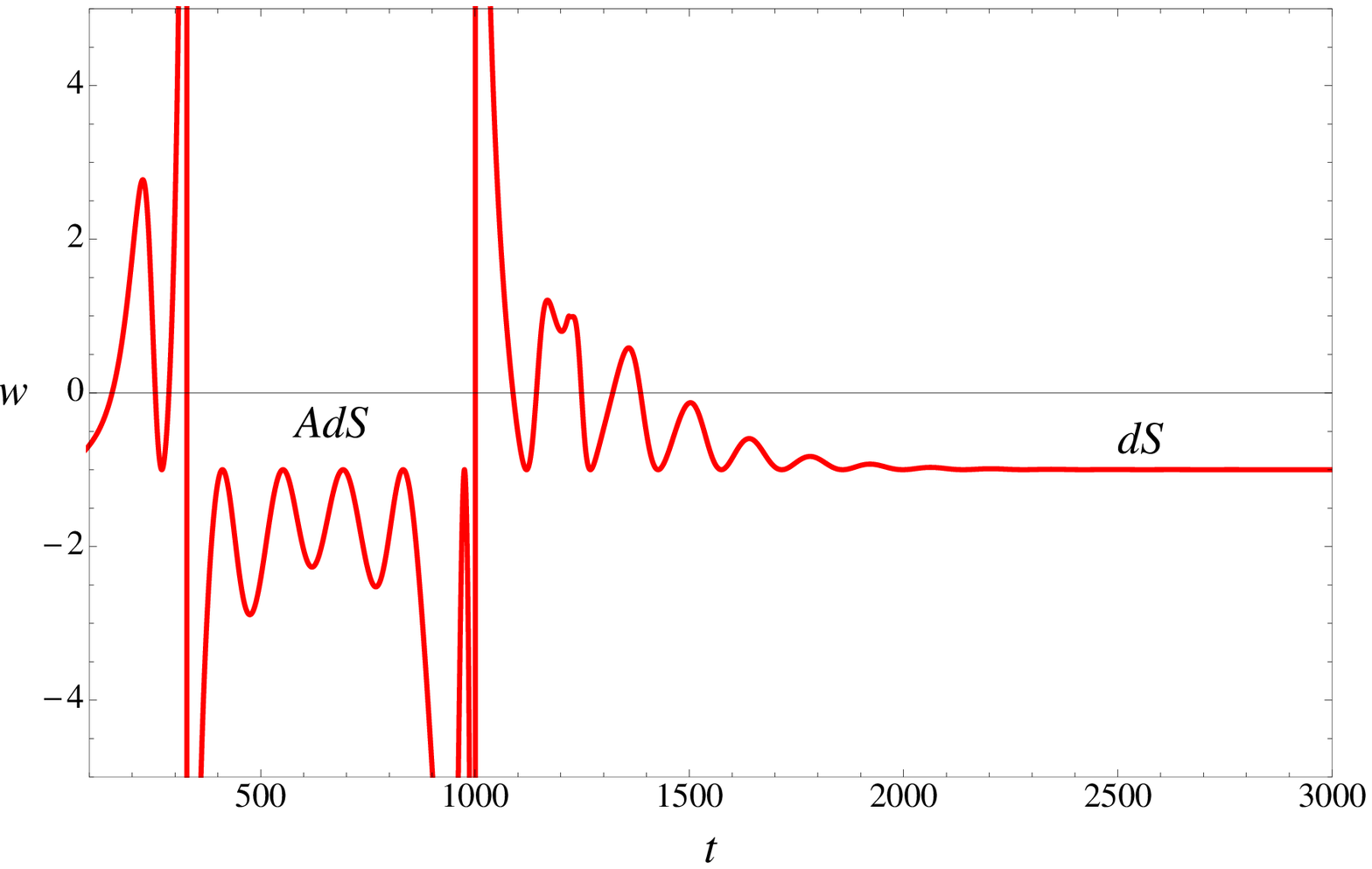}
    \label{fig:wdw1}
    }
 \caption{Figures (a) and (b) respectively show the evolution
 of the potential and the equation of state of the scalar field corresponding to \fref{fig:doublewell1}. The initial conditions for these plots are:
 $a(0)=10.5,~\phi(0)=-1.4,~\dot\phi(0)=0.14$ (in Planck units). The parameters of the
 potential are taken to be: $\alpha=0.28,~\beta=0.16, ~\delta=10$ and $V_o = 10^{-6}$.}
  \label{fig:doublewell1pot}
\end{figure}

In the evolution with a double well potential, we want to understand the way
transition from the AdS to dS vacuum takes place without collapsing into the big-crunch
singularity. In this subsubsection, we give the initial conditions in the expanding phase of the bubble universe when the field is high up in the
potential so that the field is not in the AdS vacuum in the beginning, i.e. the field is near the
point A shown in the \fref{fig:potential}. The evolution of scalar field and the scale factor is shown in \fref{fig:doublewell1}. The solid curve in this figure depicts the LQC evolution
while the dashed curve shows the classical trajectory. We have chosen initial conditions when the spacetime curvature is very small compared to the Planck scale. Thus, initially there is little difference between the classical and LQC dynamics.
As the evolution takes place, the field rolls
down the potential and crosses point `B'. During this process, the field loses its kinetic energy due to Hubble
friction as the scale factor expands. As the field enters the negative part of the potential near
the AdS vacuum, it has a very small kinetic energy, so it spends some time at the bottom of
the potential, corresponding to point `C'.
During this phase, as shown in \fref{fig:phidw1}, the
scalar field slowly oscillates in the AdS well around the bottom of the potential `C'.
There, the negative potential energy dominates the kinetic energy and the total energy density
of the field remains almost constant, as shown in \fref{fig:potentialdw1}, where the potential is
almost constant negative as long as the field remains close to the bottom of the potential. The
equation of state of the scalar field is shown in the \fref{fig:wdw1}. As long as the value of
Hubble friction is small enough, the scalar field dwells around the AdS vacuum for a finite
duration.
Due to the negative energy density, the scale factor reaches a maximum size and starts to contract, making the Hubble rate negative.  After the re-collapse takes place (shown in
\fref{fig:adw1}), the kinetic energy of the field increases  due to anti-friction originating
from the negative Hubble rate. As a result, the field starts to roll up with a high kinetic energy.
In the further evolution, as soon as the energy density of the field approaches Planckian value, the
quantum geometric effects come into dominance,  and the departures between classical theory and LQC become significant. The universe in LQC undergoes a non-singular
bounce, as shown in \fref{fig:adw1}. This behavior stands in a sharp contrast to the classical theory where a big crunch singularity is reached. As
shown in \fref{fig:doublewell1}, the scale factor in the classical theory (dashed curve), goes to
big crunch, whereas the LQC trajectory bounces. The kinetic energy of the scalar field in the
classical theory, diverges due to diverging Hubble rate. Due to this, the
value of the scalar field diverges as well, and the scalar field does not settle into the dS
vacuum.

In LQC, following the bounce, the Hubble rate again becomes positive, and its value
increases rapidly.
If the kinetic energy is high enough, the scalar field shoots into the positive well of
the potential, as shown in the \fref{fig:doublewell1}. In this phase of the evolution, the scale
factor expands due to the positive Hubble rate and the field slows down, this time getting
trapped into dS vacuum. In the dS vacuum state, the scale factor undergoes exponential
expansion and the due to high Hubble friction the field remains in the dS vacuum, undergoing
exponential inflation during the rest of the evolution.

\begin{figure}
    \includegraphics[width=0.5\textwidth]{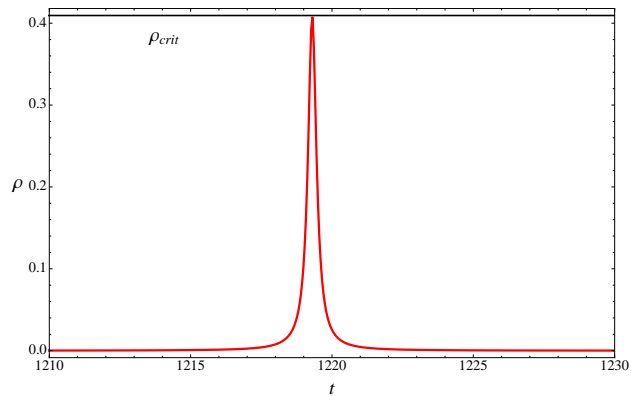}
    \caption{This figure shows the evolution of the energy density of the scalar field,
    corresponding to \fref{fig:doublewell1}, as the
    bounce takes place. The plot is zoomed in near the bounce in order to show the smooth
    evolution. It is clear to see that unlike in the classical theory where the energy density
    would have diverged to infinity, in LQC it remains finite throughout the evolution. The value
    of energy density shown in this figure is in the units of Planck density $\rho_{\rm Pl}$}
    \label{fig:rhodw}
\end{figure}
\begin{figure}
     \includegraphics[width=0.5\textwidth]{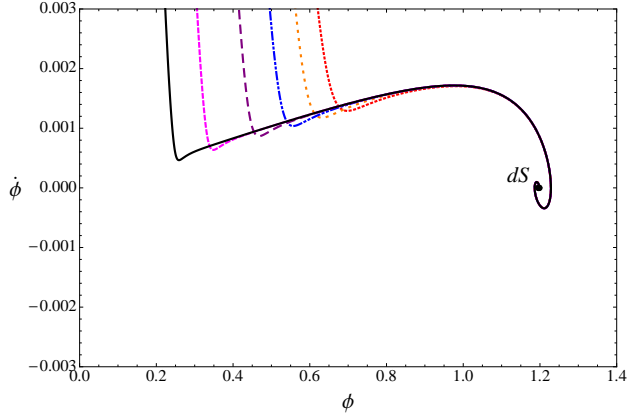}
     \caption{This figure shows the dynamical phase trajectories of the scalar field at late
     times when the field climbs up to the positive part of the potential. Different trajectories
     correspond to different initial conditions on the value of the scalar field velocity,
     $\dot \phi$. It is clear from the plot that the terminal dS phase is obtained for a range
     of initial condition. This implies that dS vacuum is a future attractor for such a choice of initial
     conditions. The parameters of the potential are taken to be: $\alpha=0.38,~\beta=0.62, \delta=10$ and $V_o = 10^{-6}$.}
     \label{fig:phasedw}
\end{figure}

Thus, it is clear from the above discussion and \fref{fig:doublewell1} that in the effective
dynamical description of LQC, there is a non-singular and smooth transition from AdS to dS
vacuum. It is also important to note that the energy density of the scalar field remains finite
throughout the evolution, and scale factor remains greater than Planck length at the bounce.
\fref{fig:rhodw} shows the evolution of energy density of the scalar field close to the bounce as
the transition from AdS to dS vacuum takes place. The horizontal curve denoted by
$\rho_{\rm crit}$ shows the upper bound on the energy density, given as
$\rho_{\rm crit}\approx0.41\rho_{\rm Pl}$. It is evident from the figure that, unlike in the
classical theory where the energy density diverges, in LQC the energy density
remains bounded. This is a distinguished feature of LQC, that all the curvature scalars always
remain finite during the evolution.

It is worth mentioning that, the initial conditions are a little fine tuned, in order for
the field to roll down and spend some time at the bottom of the AdS vacuum. However,
the behavior described above remains qualitatively similar for a small perturbation around the initial conditions. As
shown in \fref{fig:phasedw}, it also turns out that the dS vacuum phase is a future attractor for
all such initial conditions for which the evolution takes place according to \fref{fig:doublewell1}.
It is also evident that the dynamical trajectories corresponding to different initial
conditions tend to the dS vacuum in their future evolution.

\begin{figure}[tbh!]
     \subfigure[]
      {
        \includegraphics[width=0.47\textwidth]{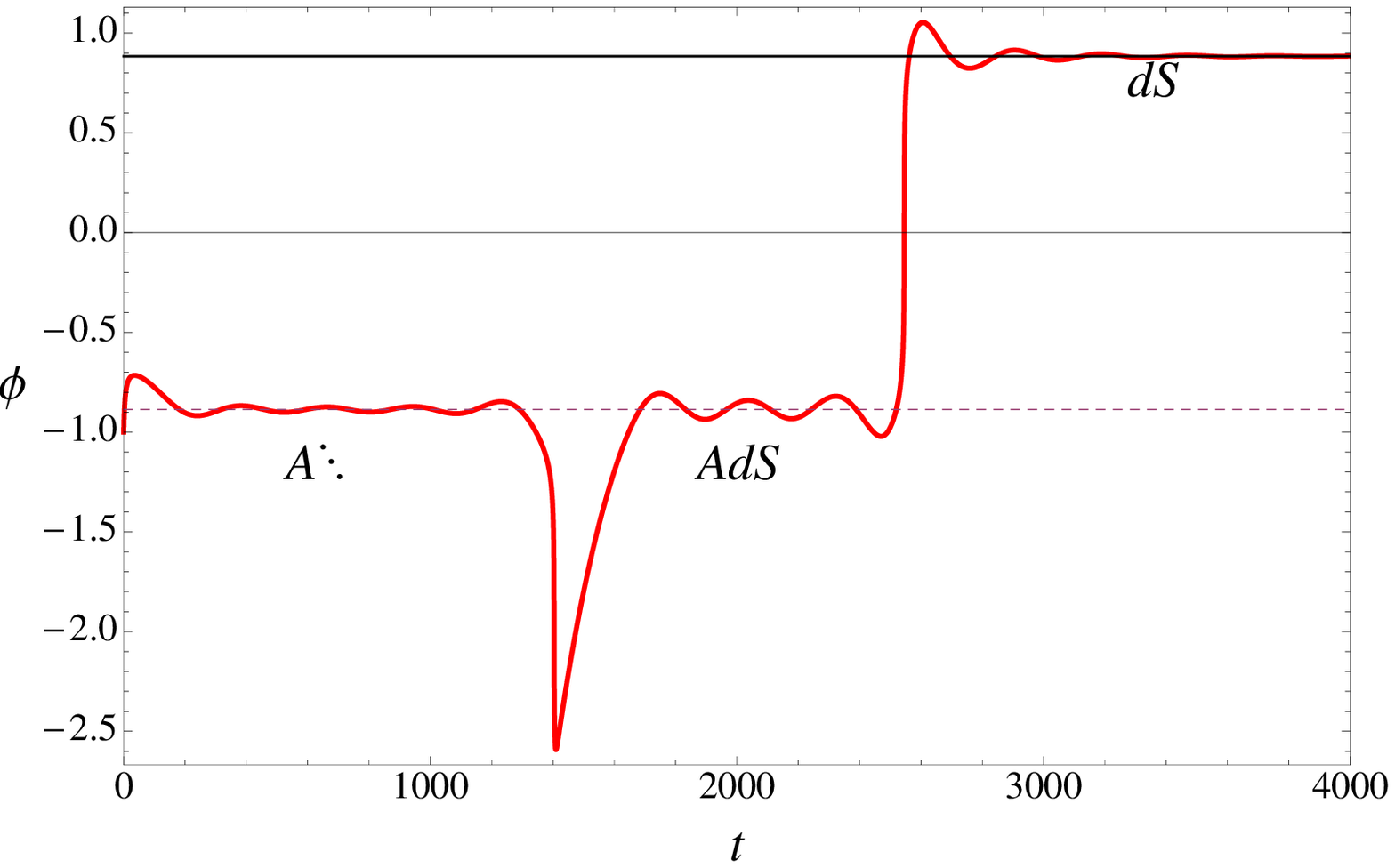}
        \label{fig:2adsphidw}
       }
     \subfigure[]
      {
        \includegraphics[width=0.47\textwidth]{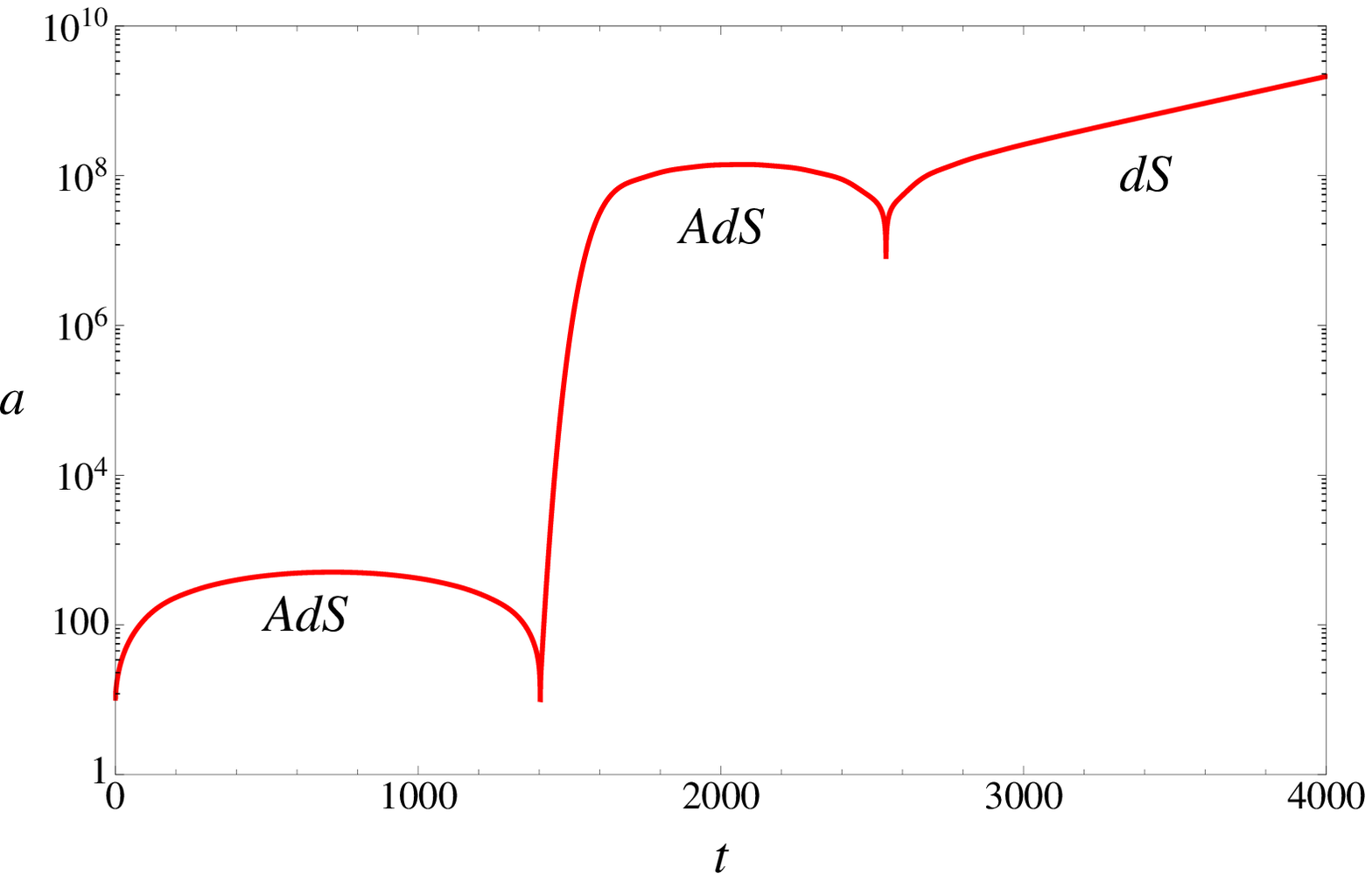}
        \label{fig:2adsadw}
       }
   \caption{This figure shows the an example of evolution when the field spend more time in the AdS well while the scale factor undergoes two cycles of recollapse and bounce. After these two cycles, the field makes a transition to $dS$ vacuum. The initial conditions for these plot are:
 $a(0)=10.5,~\phi(0)=-1.4,~\dot\phi(0)=0.07$ (in Planck units). Parameters of the potential are taken to be the same as in \fref{fig:doublewell1}. }
   \label{fig:2adsdoublewell}
\end{figure}

Depending on the initial conditions, it is possible that the scalar field may go through multiple AdS phases before making a non-singular transition to the dS vacuum.
\fref{fig:2adsdoublewell} shows such an interesting phenomenological case for the double well
landscape potential. As compared to the evolution shown in \fref{fig:doublewell1}, the
scale factor undergoes two cycles of recollapse while the scalar field oscillates in the AdS
well. It turns out that after the first bounce, the velocity of the scalar field, $\dot\phi$ is
negative. As a result, instead of crossing the barrier between AdS and dS vacua, the field
attempts to climb back towards `A' but fails due to infinite barrier on that side. However,
during the second cycle, the field has a positive velocity with enough kinetic energy to
transit to the dS vacuum. Similarly, the initial conditions can be fine tuned to obtained more
cycles of recollapse before transiting to dS vacuum. By tuning the parameters of the potential,
one may also obtain initial conditions such that the field never shoots off to the positive part
of the potential and keeps oscillating around the AdS vacuum. In this way the evolution of
spacetime under the landscape potential in LQC has a rich phenomenology.

\subsubsection{Evolution in double well potential: initial conditions in the AdS well}
\begin{figure}[tbh!]
     \subfigure[]
      {
        \includegraphics[width=0.47\textwidth]{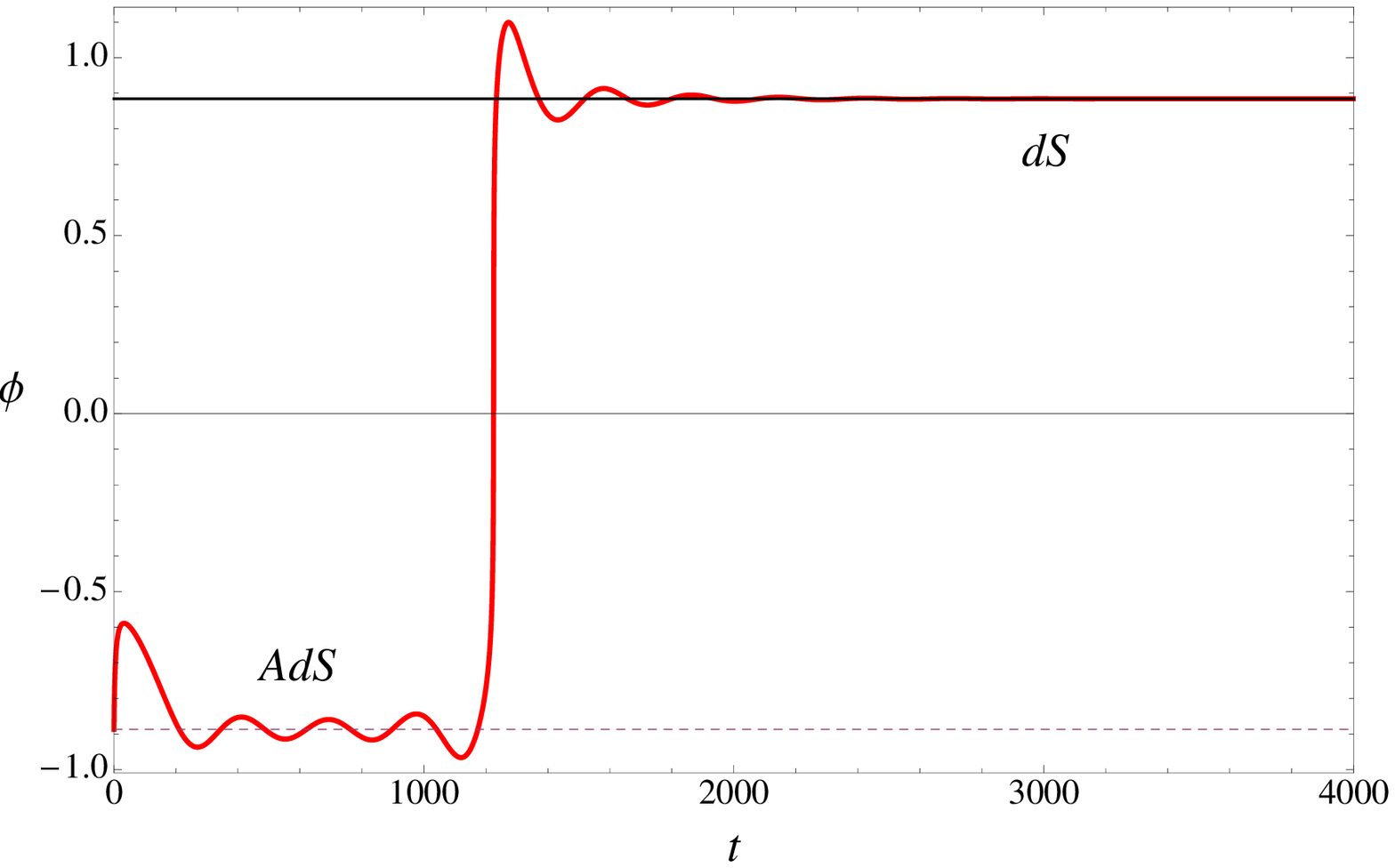}
        \label{fig:adsphidwC}
       }
     \subfigure[]
      {
        \includegraphics[width=0.47\textwidth]{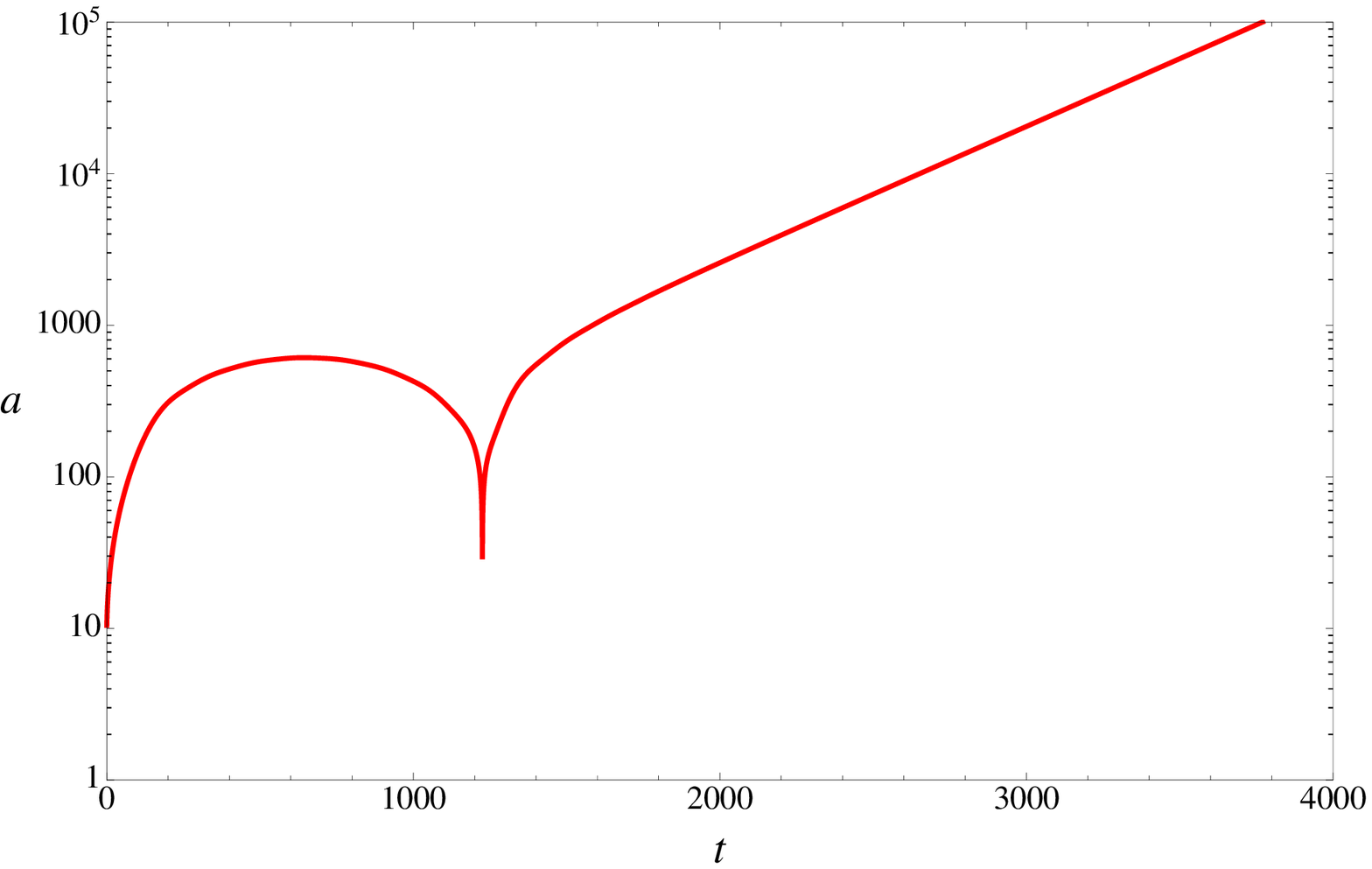}
        \label{fig:adsadwC}
       }
   \caption{This figure shows the an example of evolution when the initial conditions are provided at the minimum of the AdS well corresponding to the point `C' in \fref{fig:potential}. It is clear to see that the field undergoes a non-singular evolution, and ends up in the dS vacuum. During this evolution the scale factor undergoes one cycle of recollapse in the AdS phase. The initial conditions for these plot are:  $a(0)=10.5,~\phi(0)=-1.4,~\dot\phi(0)=0.17$
(in Planck units). Parameters of the potential are taken to be the same as in \fref{fig:doublewell1}.}
   \label{fig:adsdoublewellC}
\end{figure}

So far we have discussed the evolution of the field in a double well potential (as shown
\fref{fig:potential}) when the initial conditions are given near the point `A'. That is, the field
begins to roll down from a positive part of the potential, and approaches the AdS well. Let us now
consider initial conditions so that the field starts its evolution from the bottom of the AdS well
with a very small kinetic energy. \fref{fig:adsdoublewellC} shows the evolution of the scalar
field and the corresponding scale factor. Since the field is in the AdS vacua, future evolution leads to a recollapse, and the
scale factor starts decreasing. In subsequent evolution,  the curvature of spacetime becomes
Planckian, and due to the quantum geometric effects, the scale factor goes
through a non-singular bounce. If the kinetic energy is sufficient and the field has the  correct sign
of $\dot\phi$ (positive in this case), it settles in the dS well (corresponding to point `D'). In the further evolution,
as shown in the \fref{fig:adsdoublewellC}, as the field settles in the dS vacuum, while the scale
factor undergoes an exponentially expanding phase. Thus we obtain a non-singular transition from AdS to dS phase.

\subsection{Triple well potential}
\begin{figure}
    \includegraphics[width=0.5\textwidth]{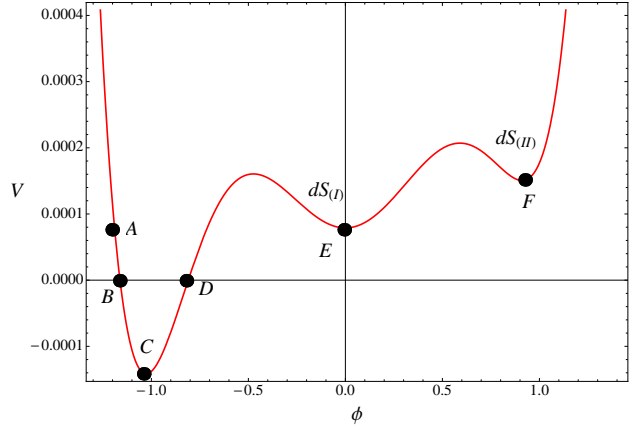}
\caption{This figure shows an example of a triple well landscape. The point `A' refers to the position of the scalar field for a class of initial conditions when the field begins its evolution out the of the AdS well. `B' and `D' denote the zeros of the potential, and `C' marks the AdS vacuum. Points `E' and `F' correspond to the de-Sitter vacua dS$_{(I)}$ and dS$_{(II)}$ respectively.}
\label{fig:potentialtriple}
\end{figure}

We now consider a triple well potential in the landscape scenario which has one AdS vacuum and two
different dS vacua. We consider the potential of the following form:
\be
\label{eq:pottriple}V(\phi)= V_o \left(\left(\f{\phi^2}{\a^2}-\delta\right)^3 + \nu \left(\f{\phi^2}{\a^2}-\delta\right)^2 + \f{\b}{\a}\phi + \Omega\right)
\ee
where $\alpha,\, \beta,\, \nu,\, \delta,\, \Omega$ and $V_o$ are parameters of the potential.
The landscape potential given by the above equation has the shape as shown in the
\fref{fig:potentialtriple}. In this figure, point `A' refers to the position of the scalar field for a class
of initial conditions when the field begins its evolution away from the AdS well. Points `B' and `D'
denote the zeros of the potential, `C' marks the AdS vacuum, and `E' and `F' correspond to the
de-Sitter vacua dS$_{\rm (I)}$ and dS$_{\rm (II)}$ respectively. The arrow in \fref{fig:potentialtriple}
depicts the case of the initial condition when the field is considered to be rolling down, and it enters the AdS well in future
evolution. As compared to the double well landscape potential discussed in the previous
subsection, there are two possible end states in the present case. That is, after the field comes
out of the AdS well, it can either settle in the first dS vacuum (dS$_{\rm (I)}$) or in the second
dS vacuum (dS$_{\rm (II)}$). Whether the field ends up in dS$_{\rm (I)}$ or dS$_{\rm (II)}$
depends on the magnitude of the potential and the initial velocity of the scalar field. We will now
solve the equations of motion for the landscape potential given in \eref{eq:pottriple}, and obtain
transitions to both the dS vacua, namely dS$_{\rm (I)}$ and dS$_{\rm (II)}$.

The initial evolution of the scalar field is similar to that in the case of double well potential.
As the field rolls down to the AdS well, the scale factor begins to contract. In the further
evolution, instead of going into big-crunch, the scale factor bounces from a finite non-zero
value. During the bounce, the field gains a lot of kinetic energy owing to
the anti-friction caused by negative Hubble rate. After the bounce, the field may either roll back
in to the AdS well (which happens if the velocity of the scalar field becomes negative), or climb
up the potential all the way to the positive hills (dS vacua). From this moment onwards, if the field has
enough kinetic energy, it passes through the dS$_{\rm (I)}$ vacuum and sits in the
dS$_{\rm (II)}$ vacuum. Otherwise, the field will settle down in the first de-Sitter well, i.e.
dS$_{\rm (I)}$.  In the following subsubsections, we consider two types of initial conditions: (i)
when the scalar field begins to roll down from the positive part of the potential
(corresponding to point `A' in \fref{fig:potentialtriple}) on the left of the AdS vacuum, and (ii)
when the scalar field starts its evolution from the bottom of the AdS well (corresponding to point
`C' in \fref{fig:potentialtriple}).

\begin{figure}[tbh!]
     \includegraphics[width=0.5\textwidth]{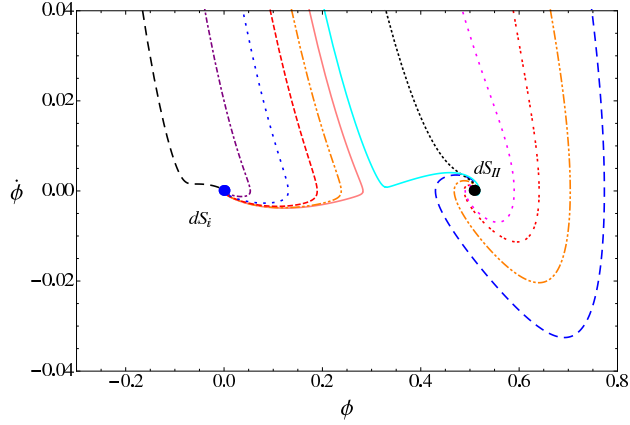}
     \caption{This figure shows the dynamical phase trajectories of the scalar field at late
     times when the field climbs up to the positive part of the potential. Different trajectories
     correspond to different initial conditions on the value of the scalar field velocity,
     $\dot \phi$. It is evident from the figure that there are two possible future attractors for
     these class of solutions. Some of the trajectories end up into first dS vacuum (denoted by
     dS$_{\rm (I)}$) and others into second dS vacuum (denoted by dS$_{\rm (II)}$). The parameters of the potential are taken to be: $\alpha=0.16$, $\beta=129.6$, $\delta=2.9$, $\nu=-13$, $\Omega=180$ and $V_o = 10^{-6}$. }
     \label{fig:phasetw}
\end{figure}
\begin{figure}[tbh!]
     \subfigure[]
      {
        \includegraphics[width=0.47\textwidth]{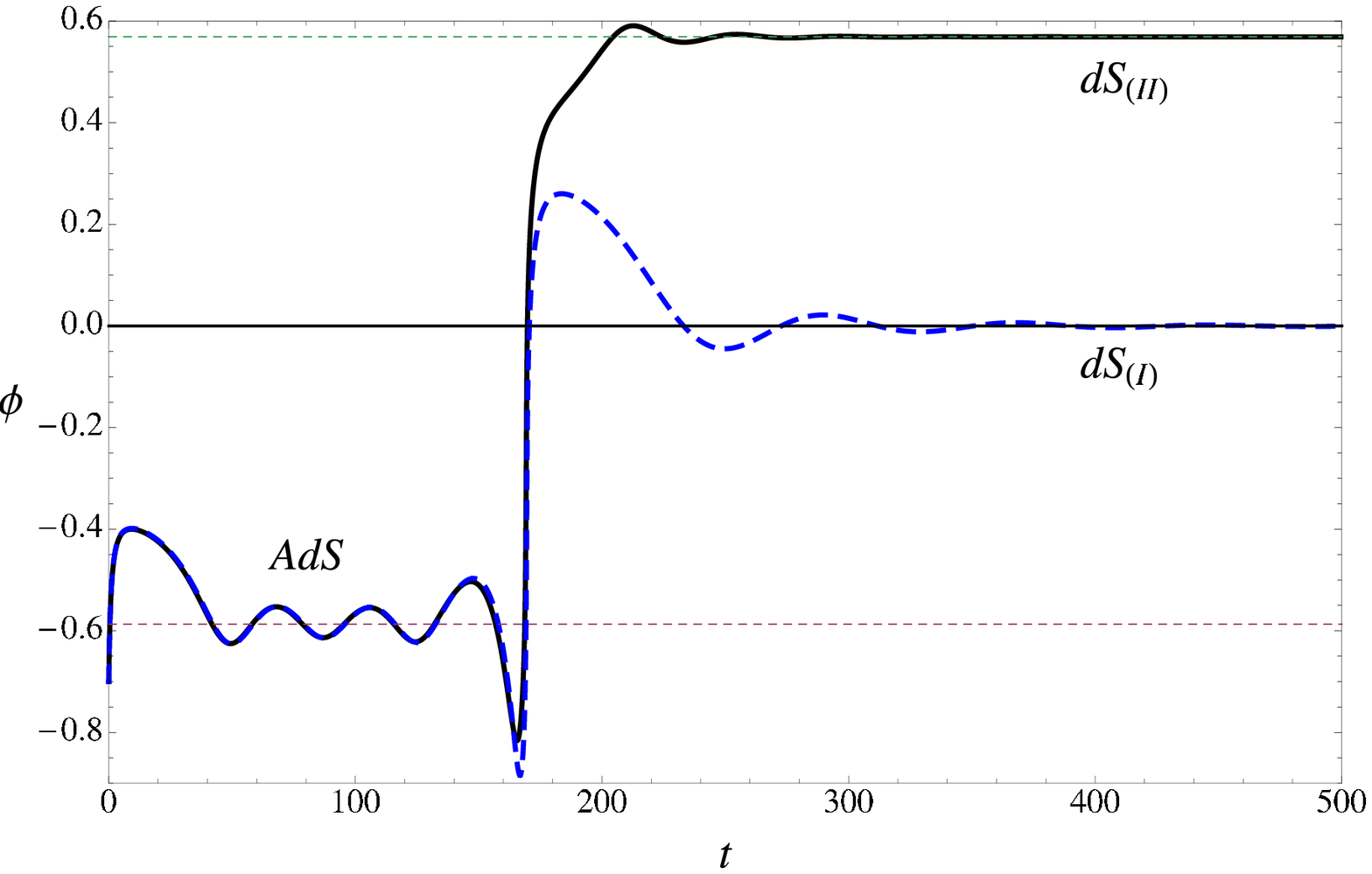}
        \label{fig:phitw}
       }
     \subfigure[]
      {
        \includegraphics[width=0.47\textwidth]{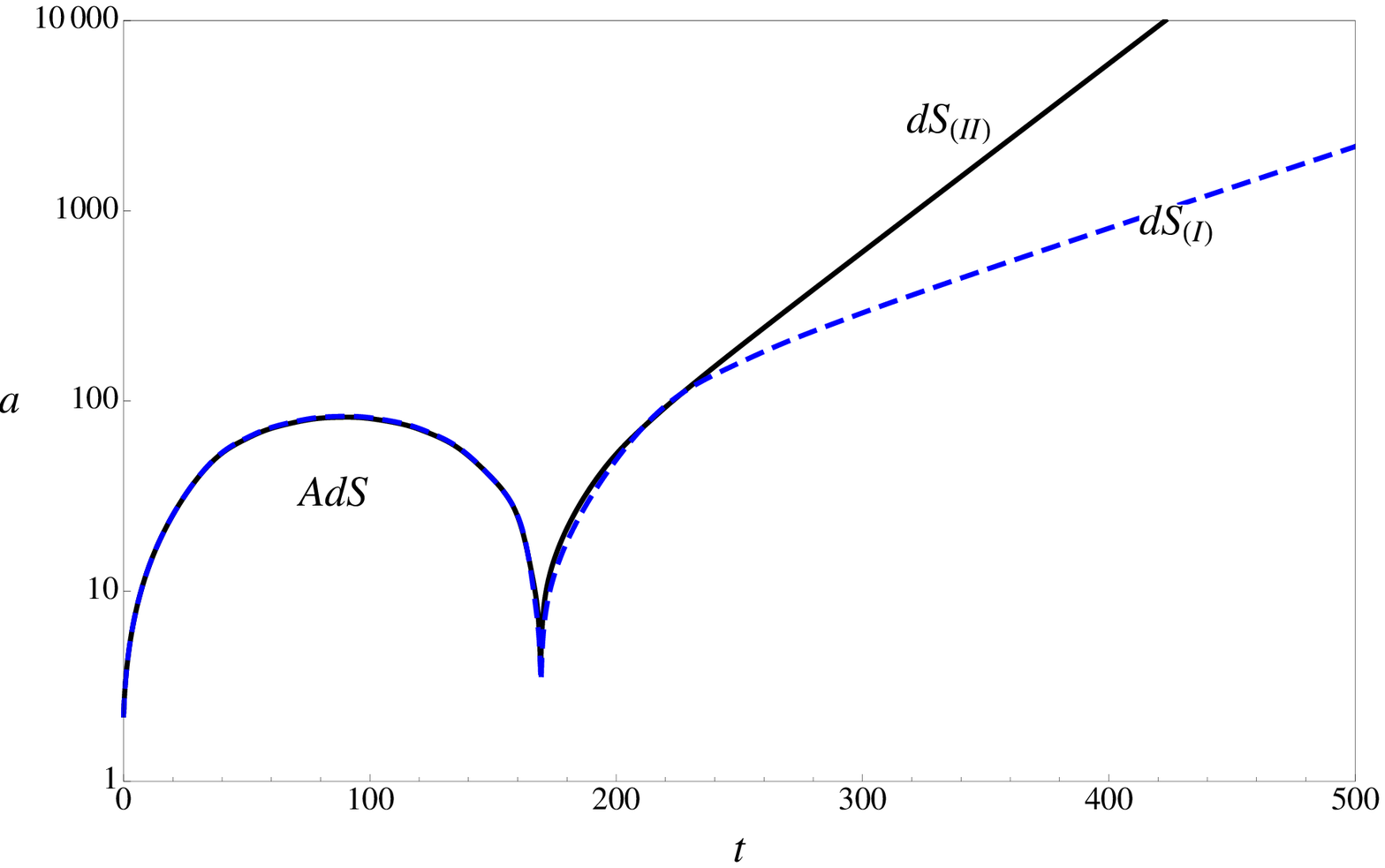}
        \label{fig:atw}
       }
   \caption{This figure shows two examples of evolutions in triple well landscape potential. The solid (black) curve corresponds to the transition from $AdS$ to dS$_{\rm (II)}$ and the dashed (blue) curve shows the transition from $AdS$ to dS$_{\rm (I)}$. The figure (a) shows the evolution of the scalar field and (b) shows the corresponding scale factor which undergoes a bounce while the field makes a transition from AdS to one of the dS vacua. It is evident from the figure that in the beginning of the evolution, the scalar field spends some time in the AdS vacuum and then depending on the initial conditions and the parameters of the potential there is a non-singular transition from the AdS vacuum to one of the false dS vacua. The initial conditions for the evolution shown in these plots are:  $a(0)=7,~\phi(0)=-0.75,~\dot\phi(0)=1.17\times10^{-2}$ (in Planck units) for the solid curve, and $a(0)=7,~\phi(0)=-0.75,~\dot\phi(0)=1.18\times10^{-2}$ for the dashed curve. The parameters of the potential are: $\alpha=0.16$, $\beta=75.78$, $\delta=4.25$, $\nu=-13$, $\Omega=180$ and $V_o = 10^{-6}$.}
   \label{fig:triplewell}
\end{figure}

\subsubsection{Evolution in triple well potential: initial conditions at `A'}

\fref{fig:phasetw} shows the phase diagram of the dynamical trajectories of the scalar field,
in the triple well landscape potential, when the initial conditions are provided in the positive part
of the potential out of the AdS well, corresponding to point `A' in \fref{fig:potentialtriple}.
Different trajectories in this phase plot correspond to different initial values of the scalar field
velocity, $\dot\phi(0)$. It is evident that in the future evolution, some of the trajectories end up in
the first de-Sitter vacuum dS$_{\rm (I)}$ (corresponding to point `E' in \fref{fig:potentialtriple})
while others in the second vacuum dS$_{\rm (II)}$ (corresponding to point `F' in
\fref{fig:potentialtriple}). In this way, the two de-Sitter vacua are future attractors of the
corresponding dynamical trajectories. Let us now analyze the time evolution of the scalar field
and the scale factor as the field passes through the landscape potential.
\begin{figure}[tbh!]
     \subfigure[]
      {
        \includegraphics[width=0.47\textwidth]{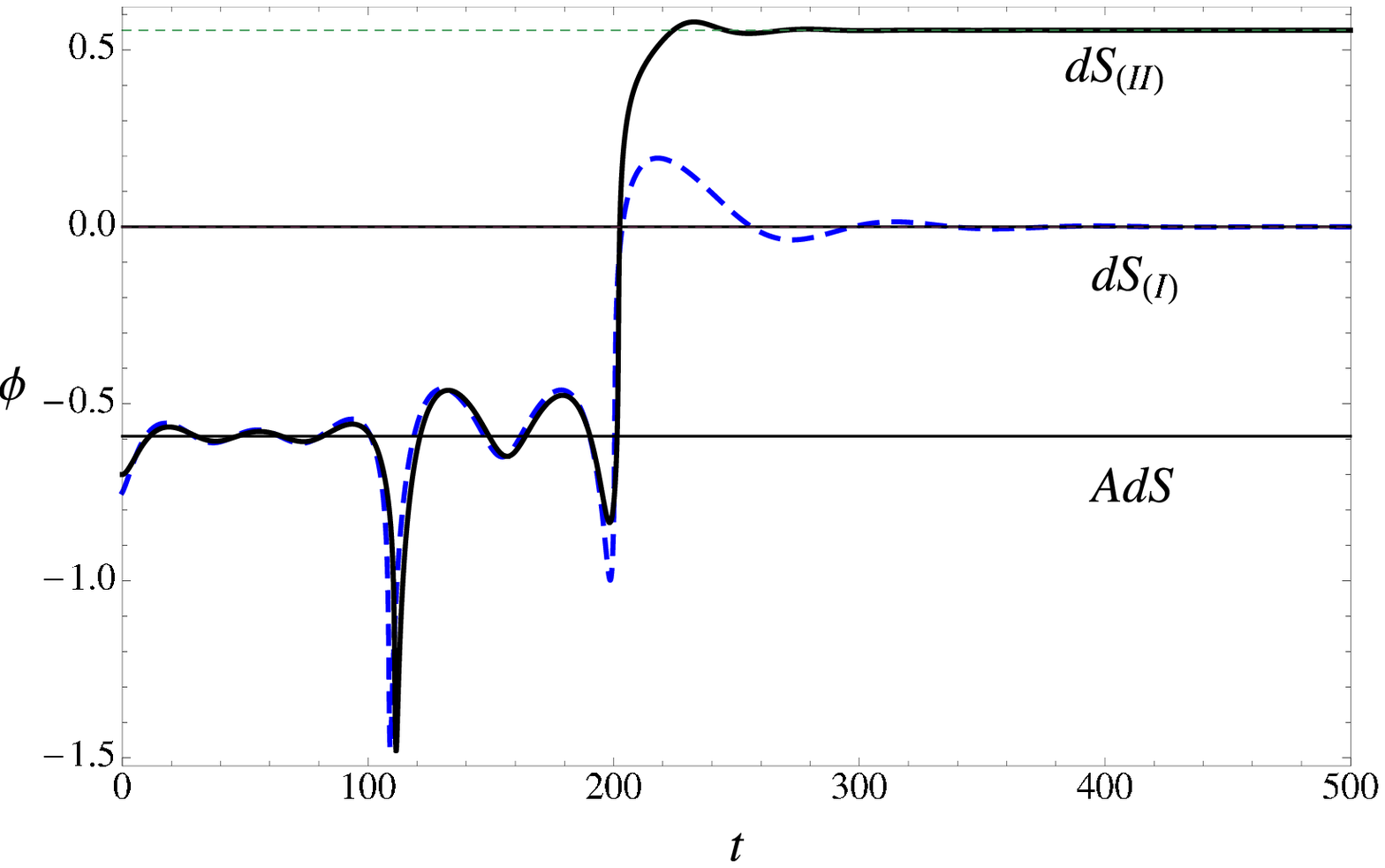}
       }
     \subfigure[]
      {
        \includegraphics[width=0.47\textwidth]{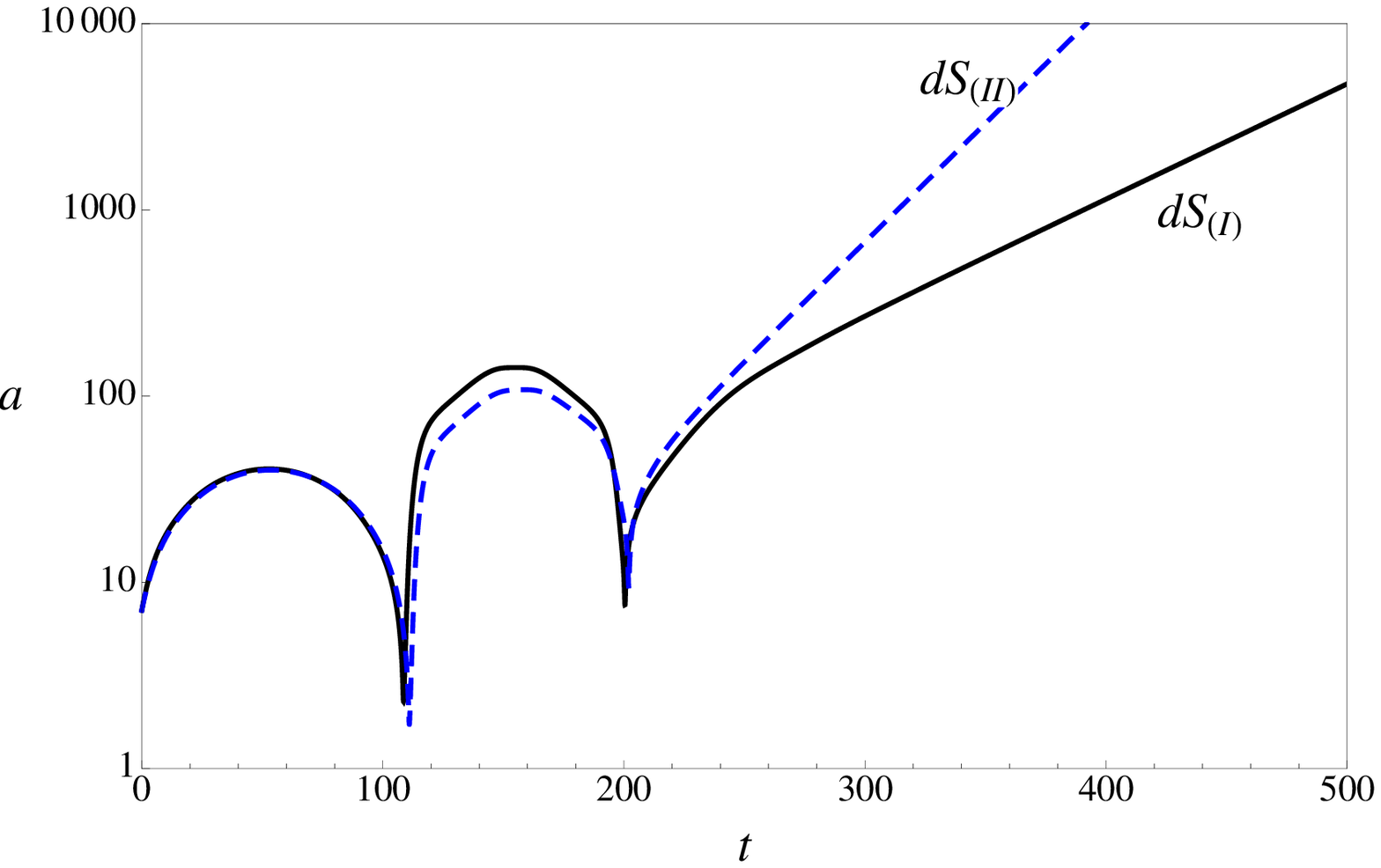}
       }
   \caption{This figure shows the an example of evolution when the field spend more time in the AdS well while the scale factor undergoes two cycles of recollapse and bounce. The dashed curve shows the transition to dS$_{\rm (I)}$, while the solid curve corresponds to the transition to dS$_{\rm (II)}$. The initial conditions for the evolution shown in these plots are:
 $a(0)=7,~\phi(0)=-0.75,~\dot\phi(0)=1.2\times10^{-2}$ (in Planck units) for the solid curve, and $a(0)=7,~\phi(0)=-0.7,~\dot\phi(0)=5.6\times10^{-4}$ (in Planck units) for the dashed curve. The parameters of the potential are taken to be the same as in \fref{fig:triplewell}.}
   \label{fig:2adstriplewell}
\end{figure}

\fref{fig:triplewell} shows the evolution of the scalar field and the corresponding scale factor
for two different situations. In one of these situations, the field makes a transition from the
AdS vacuum to the first dS vacuum (dS$_{\rm (I)}$), and in the other case the field ends up in
the second dS vacuum (dS$_{\rm (II)}$). The evolution of the corresponding scale factors
show that during these transitions, the universe undergoes a non-singular bounce instead of
falling into a big-crunch singularity. As a result, the scale factor bounces from a non-zero finite
value. During the entire evolution, the energy density and the expansion scalar remain finite,
which signals the avoidance of curvature singularity in the course of `AdS-dS$_{(i)}$' vacuum
transitions. \fref{fig:2adstriplewell} shows two examples of the evolution where the the scale
factor undergoes two cycles of AdS recollapse before the scalar field makes a transition from
the AdS vacuum to one of the dS vacua. At the end of the first cycle, the velocity of the scalar
field is negative, due to which instead of shooting off to one of the dS vacua, the field attempts
to climb back in the other direction. On the other hand, the velocity of the scalar field at the
end of the second cycle becomes positive and it climbs up to one of the dS vacua.
The solid curve corresponds to the transitions to dS$_{\rm (II)}$ and the dashed curves show
the evolution when the end state of the field is dS$_{\rm (I)}$. In this way, by tuning the initial
conditions one can obtain more than one AdS cycles before making the transition to dS
vacua.

\begin{figure}[tbh!]
     \subfigure[]
      {
        \includegraphics[width=0.47\textwidth]{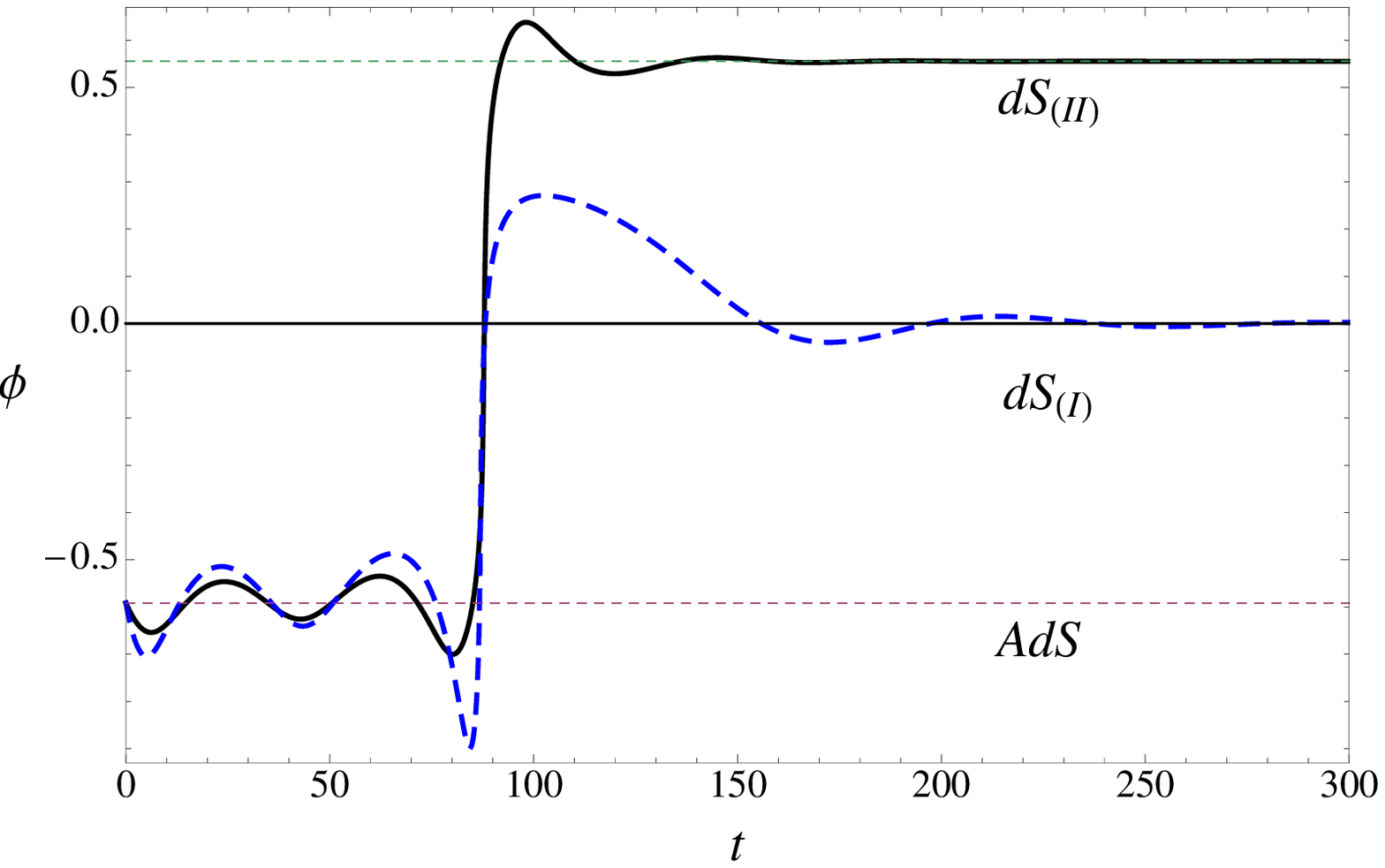}
       }
     \subfigure[]
      {
        \includegraphics[width=0.47\textwidth]{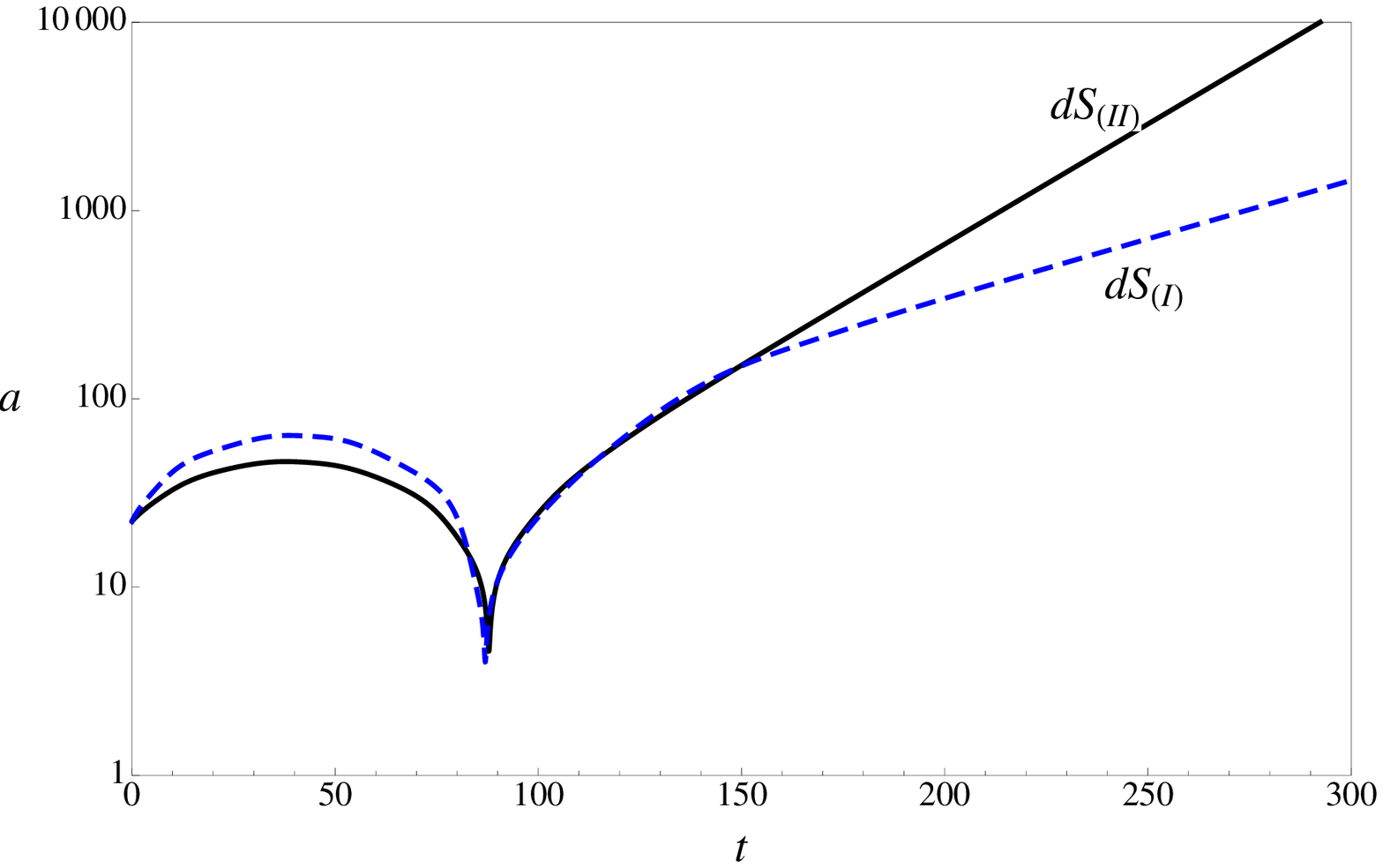}
       }
   \caption{This figure shows examples of evolution when the initial conditions are given at the bottom of the AdS well. The dashed curve shows the transition to dS$_{\rm (I)}$, while the solid curve corresponds to the transition to dS$_{\rm (II)}$. The initial conditions for the evolution shown in these plots are:
 $a(0)=22.3,~\phi(0)=-0.59,~\dot\phi(0)=2\times10^{-2}$ (in Planck units) for the solid curve, and $a(0)=22.3,~\phi(0)=-0.59,~\dot\phi(0)=4.8\times10^{-2}$ (in Planck units) for the dashed curve. The parameters of the potential are taken to be the same as in \fref{fig:triplewell}.}
   \label{fig:triplewelladsini}
\end{figure}

\subsubsection{Evolution in triple well potential: initial condition in the AdS well}
Let us now consider initial conditions so that the scalar field starts its evolution at the bottom of
the AdS well (corresponding to `C' in \fref{fig:potentialtriple}), with a small kinetic energy.
\fref{fig:triplewelladsini} shows the evolution of the scalar field and the corresponding scale
factors. The dashed curve corresponds to the transition to dS$_{(I)}$ (corresponding to point `E'
in \fref{fig:potentialtriple}), while the solid curve shows the transition to dS$_{(II)}$
(corresponding to point `F' in \fref{fig:potentialtriple}). Since, the field is inside the AdS well, in the future evolution 
there is a recollapse of the scale factor. As the scale factor decreases, the
curvature of spacetime becomes Planckian and quantum geometric effects become prominent.
As a result, instead of going into a big-crunch singularity, the scale factor undergoes a smooth
non-singular bounce  while the scalar field makes transition to one of the dS vacua.

\vskip0.5cm
Let us summarize the results of this section. We have considered a double well and a triple well potential in the effective spacetime description of the $k=-1$ model in LQC.
We have considered two types of initial conditions:
first, when the potential of the landscape takes a positive value starting out of the AdS well
and, second, when the field begins its evolution at the bottom of the AdS well. By numerically solving effective dynamical equations we show that unlike in the classical theory, there is a smooth non-singular transition
from AdS vacuum to dS vacua in LQC. Depending on the initial conditions, it is
possible that the field gets trapped inside the AdS well and
oscillates for more than one cycle of AdS recollapse before making transition to a dS vacuum. The big crunch singularity of the classical theory is generically resolved and replaced by a big bounce. This result can now be
generalized to any such landscape with more than one well. As discussed above, the LQC
effects are visible only when the energy density becomes Planckian. Such a situation arises
whenever the scale factor undergoes re-collapse (in our case while passing through the AdS
phase). So, in any generic landscape, the LQC effects will always resolve the big-crunch
singularity, as expected from the proof of resolution of all strong singularities in isotropic LQC \cite{ps09}.

\section{discussion}

In a multiverse permitting eternal inflation, a challenging problem is to understand the way a bubble universe  makes a transition from AdS phase to the
dS phase. The metric inside such bubbles is that of an open FRW model. It is straightforward to see using classical theory that in the future evolution, an AdS bubble encounters a big crunch singularity. Transitions from AdS to dS are thus not possible in GR. It has been hoped that incorporation of quantum gravity effects can provide insights to this problem. Apart from understanding the global structure of multiverse, an answer to this  question is also important to understand the measure problem in the multiverse.  Recently, a local measure proposal based on a `watcher' has been put forward \cite{Garriga:2012bc}, where probabilities are assigned to various events by an eternal observer, on a time-like geodesic, conjectured to make infinite number of transitions in the multiverse. Since the spacetime, in an AdS phase, ends with a big crunch singularity, it becomes necessary to find mechanisms to resolve singularities in the landscape scenario, so that the watcher can safely evolve from one vacua to another.

In this article, we addressed above problem  using the effective spacetime description of
LQC, which takes into account the corrections due to the quantum geometric nature of
spacetime, as understood in loop quantum gravity. Due to these quantum geometric corrections, the Friedmann
and Raychaudhuri equations are modified and the classical singularity is resolved.
In the effective 4-dimensional picture, we consider an open FRW universe with a scalar field, with two types of self
interacting potentials: a double well and a triple well potential. The form of the potentials is motivated from the landscape scenario.
The problem we are interested in is to understand the transitions from the AdS to dS vacuum in the multiverse. In
order to study the resolution of singularity and a smooth transition from AdS phase to a dS
phase we consider potentials which have at least one AdS vacuum.  With the help of explicit
numerical simulations we show that, unlike in the classical theory, the big-crunch singularity is
resolved in LQC and a smooth non-singular transition from AdS to dS phase can be obtained. Quantum gravitational
resolution of big-crunch singularity changes the global structure of multiverse, as well
as provides the watcher a safe passage from AdS to dS vacuum.
 It is to be noted that the quantum gravitational effects
come into play only when the energy density is Planckian. Far away from bounce, when the
curvature is very small and the energy density is less than a percent of the maximum allowed
energy density, the quantum geometric effects are too feeble to make any difference between
the effective trajectory of LQC and the classical theory.

In the numerical simulations of the landscape potentials, we considered two types of initial
conditions, first when the scalar field begin to roll down from a positive part of the potential,
and second when the field is already inside the AdS well. In the first type of initial conditions
the field enters the AdS well in future evolution. In both the cases, the numerical simulations
show that there is a non-singular smooth transition from AdS to dS phase, if the field
builds up sufficient kinetic energy during the bounce. We have also shown examples of
simulations where it takes two AdS cycles for the field to transit to the dS phase, giving rise to
a `AdS-AdS-dS' transition.
In the triple well potential landscape we have considered one AdS vacuum with negative
potential and two dS vacua with positive potential. As compared to the double well landscape,
there are two different possible end states from the scalar field in the triple well, i.e. the two
de-Sitter phases dS$_{\rm (I)}$ and dS$_{\rm (II)}$. The numerical simulations show that, like
in the double well, there are transitions from AdS to dS phases. We have considered the
transitions to both of the dS phases in the numerical simulations presented here. It turns out
that there are initial conditions for which there are two AdS cycles before the field settles in one
of the dS phases. For these initial conditions, there are two different future attractors
corresponding to the two dS vacua, portrayed in the dynamical trajectories of the field.

Thus, we see that quantum geometric corrections in LQC resolve the big-crunch singularity
occurring inside a bubble in the AdS phase. The non-singular evolution across the the big-
crunch enables one to extend the geodesics across the big crunch. This provides the `watcher'
a safe transition from AdS to dS phase as well as opens up promising avenues to address the
measure problem in the eternal inflation scenario.

\vskip0.5cm
\noindent
{\it{Note:}} While this manuscript was being written, J. Garriga, A. Vilenkin, J. Zhang  made us aware 
of their ongoing work on similar lines which is expected to appear simultaneously \cite{alex}.

\acknowledgements
We are grateful to Alexander Vilenkin for helpful discussions and comments on the manuscript. PS 
thanks organizers of the New Frontiers in Astronomy and Cosmology conference where discussions 
with Alexander Vilenkin led to this project.  This work is supported by NSF grant PHYS1068743 and a 
grant by the John Templeton Foundation. The opinions expressed in this publication are those of the
authors and do not necessarily reflect the views of the John Templeton Foundation. BG's research
is partially supported by the Coates Scholar Research Award
of Louisiana State University.

\appendix
\section{$k=0$ FRW universe}
\label{app:k0}
The goal of this appendix is to show that for $k=0$ isotropic landscape model, though singularities are resoled, there is no AdS-dS transition.  
The metric for $k=0$ FRW universe is given as
\be
d s^2 = - d t^2 + a ^2 \left(d x^2 + d y^2 + d z^2\right)
\ee
where $a $ is the scale factor of the homogenous and isotropic metric and $t$ is the proper time. 
In the classical theory, the flat FRW universe is governed by the following Friedmann equation
\be
H^2 = \f{8 \pi G}{3} \rho.
\label{fried}
\ee
The effective Hamiltonian for $k=0$ FRW spacetime in terms of the symmetry reduced 
triad $(p)$ and the connection $(c)$ with the lapse being $N=1$ is given as \cite{aps1,aps2,aps3}
\be
\Heff=-\f{3p^{3/2}}{8\pi G\gamma^2\lambda^2}\sin^2(\bar{\mu}c)+\f{P_\phi^2}{2p^{3/2}}+V(\phi)p^{3/2}
\label{eq:heffk0}
\ee
where $\bar{\mu}=\lambda/\sqrt{p}$, $P_\phi$ is the momentum of the scalar field and $V(\phi)$
is the self interacting potential. The equations of motion of the triad and the connection can then
be derived from the Hamilton's equations of motion
\be
\dot p=\{p,~\Heff\}, \qquad \dot c=\{c,~\Heff\}.
\ee
The equations of motion hence derived give rise to modified Friedmann equations, which can 
be written in terms of the energy density of the matter field and the Hubble rate as follows
\be
H^2 = \f{8 \pi G}{3} \rho \left(1-\f{\rho}{\rho_{\rm c}}\right),
\label{friedlqc}
\ee
where $\rho$ is the energy density of the matter field, $\rho_{\rm c} = 0.41\rho_{\rm Pl}$ is the maximum 
value of the energy density at the bounce, and $H$ is the Hubble rate given via $H=\dot a/a$.
For a scalar field the energy density ($\rho$) and the pressure ($P$) of the matter field is given via
\be
\rho = \f{1}{2} \dot\phi^2 + V(\phi) \, \quad {\rm and }\quad P = \f{1}{2} \dot\phi^2 - V(\phi).
\ee
The conservation of energy momentum tensor further gives
\be
\label{eq:rhodot}\dot{\rho}=-3H\left(\rho+P\right).
\ee

In order for the field to mimic an AdS phase inside the negative part of the landscape potential, the 
velocity of the scalar field should be very small $\dot\phi\approx0$ and the potential should dominate 
to give rise to a negative total energy density, in which case $\rho\approx V(\phi)$ where $V(\phi)<0$. 
The pressure of the scalar field would then be $P\approx -V(\phi)$, making the equation 
of state of the scalar field inside the negative part of potential $w\approx-1$.
However, it is important to emphasize that in the $k=0$ model the energy density can not be negative. 
 From \eref{friedlqc} it is clear that for any physical solution one must have $H^2\geq0$,
which leads to the following inequality on the energy density
\be
\label{k0ineq} 0 \leq \rho \leq \rho_{\rm crit}.
\ee
It is now clear from the above inequality that in a homogeneous and isotropic $k=0$ 
FRW universe it is impossible to have an AdS phase, which would require $\rho<0$ 
during the evolution of a scalar field through the negative part of the potential. 
Further, it is interesting to note that unlike the case of $k=-1$ model, at $\rho = 0$, the Hubble rate in $k=0$ model in LQC vanishes. Eq.(\ref{eq:rhodot}) then implies that $\dot \rho = 0$.

Hence, the energy density can never be negative in the $k=0$ model and there can not be change in its sign. In contrast to the $k=-1$ model, this is dynamically forbidden.
\begin{figure}[tbh!]
    \subfigure[]
    {
    \includegraphics[width=0.473\textwidth]{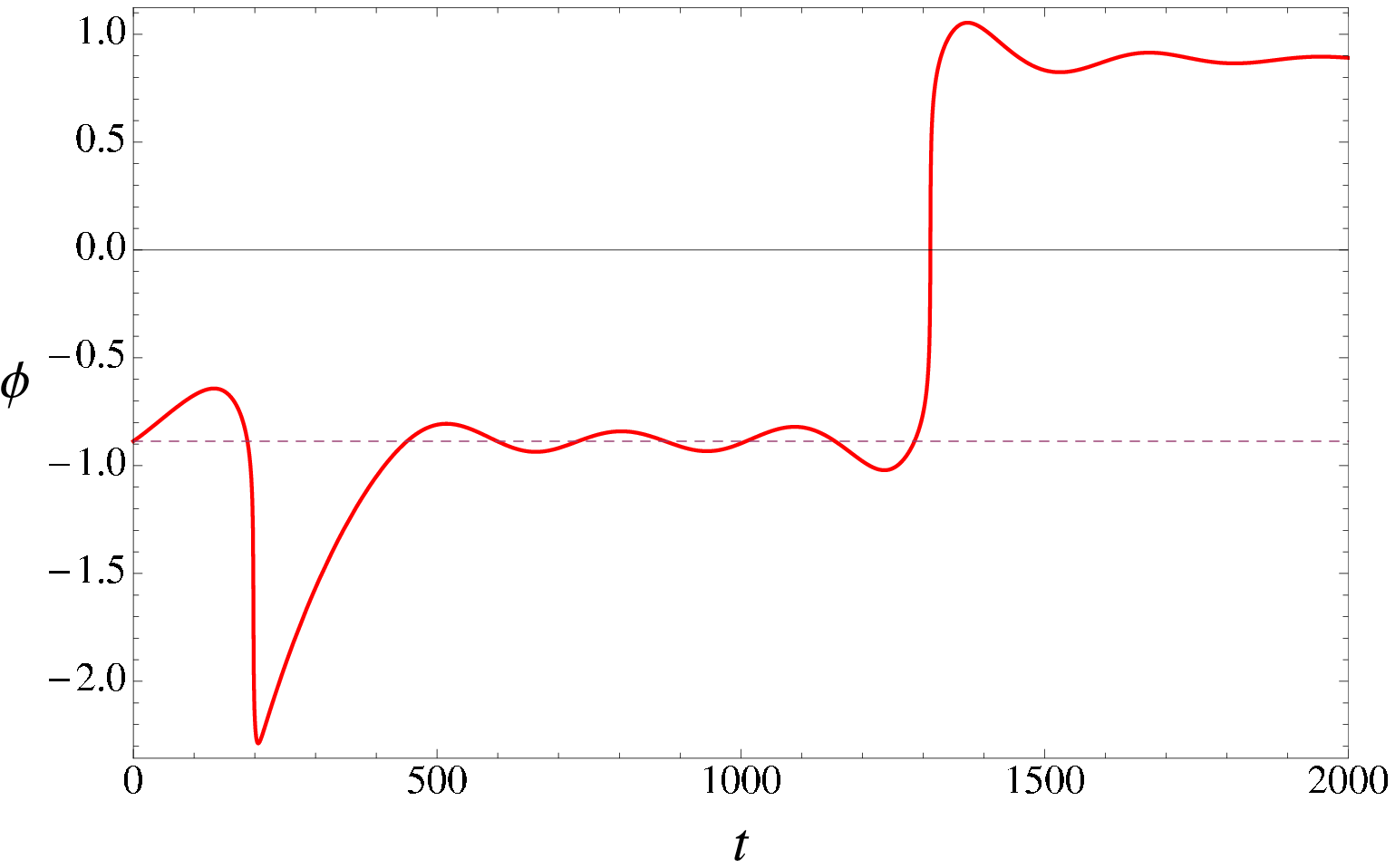}
    \label{fig:phik0}
    }
    \subfigure[]
    {
    \includegraphics[width=0.475\textwidth]{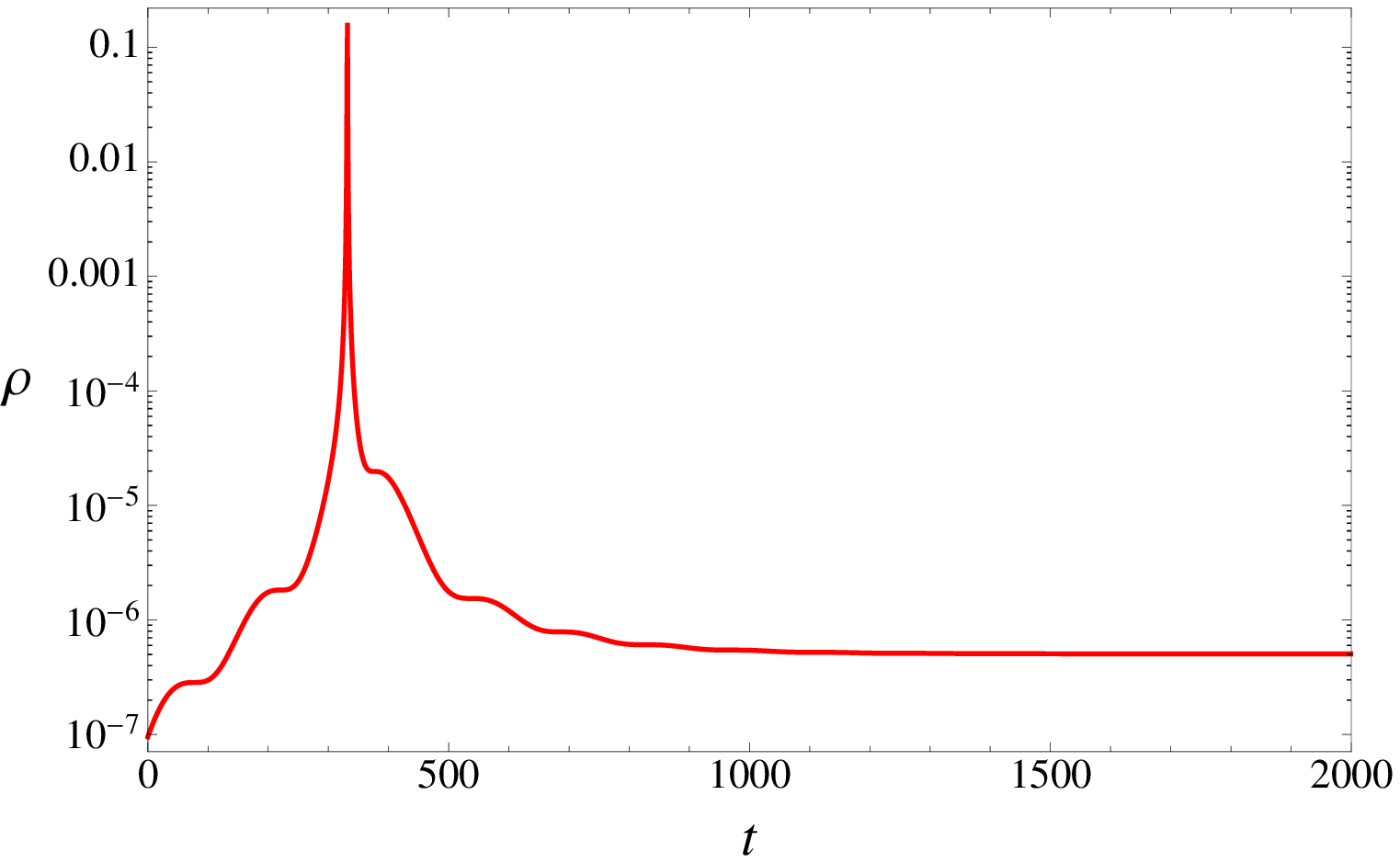}
    \label{fig:rhok0}
    }
 \caption{These figures show the evolution of a scalar field in a double well landscape 
potential (\eref{pot}) in a $k=0$ FRW universe. Figures (a) and (b) respectively show the evolution of the scalar field and the energy density. The dashed curves in Fig. (a) denotes the value of the scalar field where the potential is negative minimum. It is
clear from these figures that during the entire evolution the energy density always remains positive, even when the field evolves through the negative minimum of the potential. The initial conditions for these plots are:
 $a(0)=70.7,~\phi(0)=-0.887,~\dot\phi(0)=0.002$ (in Planck units). The parameters of the
 potential are taken to be: $\alpha=0.28,~\beta=0.16,~ \delta = 10$ and $V_o = 10^{-6}$.}
  \label{fig:k0}
\end{figure}

\fref{fig:k0} shows the evolution of the scalar field and energy density 
in a double well landscape potential given by \eref{pot}. It is evident from the figure that even though the 
scalar field evolves through the negative minimum of the potential (denoted by dashed horizontal line in 
panel (a)), the energy density remains positive through out, implying that the evolution never 
becomes AdS in a $k=0$ FRW universe.
However, it is possible to achieve $\rho<0$ and still satisfy $H^2>0$ if there is an additional 
positive term on the right hand side of \eref{fried} and \eref{friedlqc}, which is exactly the case 
in an open $k=-1$ FRW universe. Thus, the analysis of the flat FRW model
in the landscape scenario fails to capture the relevant physics of AdS-dS transitions.

\label{discussion}

\end{document}